\documentclass[twocolumn,preprintnumbers,amsmath,amssymb,floatfix,prb]{revtex4}

\usepackage{graphicx}
\usepackage{epsfig}
\usepackage{dcolumn}
\usepackage{bm}

\input epsf

\begin{document}



\title{Analogs of quantum Hall effect edge states in photonic crystals}

\author{S. Raghu}
\email{sraghu@princeton.edu}
\affiliation{Department of Physics, Princeton University,
Princeton NJ 08544-0708}

\author{F. D. M. Haldane}
\affiliation{Department of Physics, Princeton University,
Princeton NJ 08544-0708}
\date{February 25, 2006}

\begin{abstract}
``Photonic crystals'' built with time-reversal-symmetry-breaking
Faraday-effect media can exhibit ``chiral'' edge modes that
propagate unidirectionally along boundaries across which the
Faraday axis reverses.  These modes are precise analogs of the
electronic edge states of quantum Hall effect (QHE) systems, and
are also immune
to backscattering and localization by disorder.  The ``Berry curvature''
of the photonic bands plays a role analogous to that of
the magnetic field in the QHE.   Explicit calculations demonstrating the
existence of such unidirectionally-propagating photonic
edge states are presented.  
\end{abstract}


\maketitle
\section{Introduction}

The control of the flow of light using photonic band-gap (PBG) materials has
received considerable 
attention over the past decade \cite{joannopoulos}.  Moreover,
the potential for using
artificially structured ``metamaterials'', such as the recently discovered
``left-handed media'' \cite{left-handed}, has shown considerable technological promise.
In the past, significant progress has been achieved in the field of 
photonics by
making use of analogies with electronic systems.  For instance, the typical PBG material, 
a system
with a spatially varying and periodic dielectric function, was motivated by the well
known physics of electronic Bloch states; the dielectric scattering of light in periodic
media presents the same formal solutions as those for the scattering of electrons in 
periodic potentials.

Previous  photonic bandstructure calculations
have focused on the frequency dispersion of the photon bands; it has been usually been
assumed that a knowledge of the spectrum alone represents a 
complete understanding of dynamics of the system.
A primary goal of such calculations
has been the quest for a PBG material with a complete bandgap
throughout the Brillouin zone in some frequency range, which would prevent the
transmission of light with frequency in the range of the Band gap.  Both two and 
three dimensional bandstructures possessing these properties 
have now been discovered \cite{lincoln logs}.  

Recently, however, in the study of electronic systems, it has become apparent that,
even in the absence of interaction effects,
the dispersion relations  of the energy bands do not  fully characterize the semiclassical
dynamics of wavepackets, unless both spatial inversion symmetry and time-reversal symmetry
are unbroken\cite{sundaram niu}.     The additional information, which is not obtainable
from knowledge of the energy bands  $\epsilon_n(\bm k)$ alone, 
is the variation of the Berry curvature\cite{simon} 
$\mathcal F^{ab}_n(\bm k)$ = $\epsilon^{abc}\Omega_{nc}(\bm k)$, 
which is an antisymmetric tensor in $\bm k$-space, where $\bm \Omega_n(\bm k)$ is analogous
to a ``magnetic field'' (flux density) in $\bm k$-space.   The ``Berry curvature'' in
$\bm k$-space is related
to the Berry phase\cite{berry} in the same way that the Bohm-Aharonov phase of an electronic
wavepacket is related to
the magnetic flux density in real space.

 While the uniform propagation of 
wavepackets in perfectly translationally-invariant systems does not involve the Berry curvature,
the ``semiclassical'' description of the  \emph{acceleration} of wavepackets in media with
spatial inhomogeneity of lengthscales large compared to the underlying lattice spacing
is incomplete if it is not taken into account.
Recently Onoda {\it et al.}\cite{nagaosa} have pointed 
out the role of Berry curvature in photonic crystals
without inversion symmetry; while these authors characterize this as a ``Hall effect of light'',
the Hall effect in electronic systems is associated 
with broken \emph{ time-reversal symmetry}
rather than broken spatial inversion symmetry, and we 
have recently discussed\cite{raghu} some of the
at-first-sight-surprising effects that broken 
time-reversal symmetry could produce in
photonic systems.    

In the presence of non-vanishing Berry curvature, 
the usual ``semiclassical'' expression for
the group velocity of the wavepacket is supplemented by an additional ``anomalous''
contribution proportional to its acceleration and the local Berry curvature of the
Bloch band.   
(The semiclassical treatment of electron dynamics becomes ray optics
in the photonic context).   This ``anomalous velocity'' 
has played an important role in 
understanding recent experiments
on the anomalous Hall effect of ferromagnets \cite{ahe}, for example.

Perhaps the most remarkable among the ``exotic'' effects 
associated with Berry curvature, however, 
is the quantum Hall effect \cite{qhe},  which
has been the focus of intensive experimental 
and theoretical study in condensed matter
physics for over
two decades.  The physics of the quantum Hall regime and its connection
with Berry Curvature phenomena is now
well understood.  The possibility
of transcribing some of the main features 
of the quantum Hall effect to photonic systems, 
which brings into play
new possibilities in photonics, is the topic of this paper.  Specifically,
we shall concern ourselves with analogs of ``chiral'' (unidirectional) 
quantum Hall edge states
in photonic systems with broken time-reversal symmetry.

The quantum Hall effect is usually associated with two dimensional electron
systems in semiconductor heterojunctions in strong applied magnetic fields.  By
treating the plane of the heterojunction as a featureless two-dimensional (2D)
continuum, and considering
the quantum mechanical motion of electrons in the presence of a magnetic field, 
one obtains the electronic Landau levels.  
The key feature giving rise to the quantization of the Hall conductance
is the incompressibility of the electron fluid: either due to the Pauli 
principle at integer Landau level fillings, 
with the spectral gap to the next empty level given by the cyclotron frequency, 
or when a gap opens due to strong electron-electron 
interactions at fractional fillings\cite{laughlin}.

While in the experimentally-realized systems, 
the quantum Hall effect derives from a strong uniform
component of magnetic flux density normal to the 2D plane 
in which the electrons move, the integer quantum  
Hall effect can also in principle derive from the interplay of a periodic
bandstructure with a magnetic field.
Externally applied 
in periodic structures give rise 
to the celebrated Hofstadter model of Bloch bands with an
elegant fractal spectral structure depending on the rational value of 
the magnetic flux through the unit cell \cite{hofstadter} \cite{niu 2}.  
The influence of the lattice on the 
quantum Hall effect was further investigated in an important paper by 
Thouless et al (TKNN) \cite{tknn}, who discovered a
topological invariant of 2D bandstructures known as the ``Chern Number'' 
a quantity that was later interpreted in terms of Berry curvature \cite{simon}.

At first sight, it seems implausible that any of the phenomena associated with the quantum
Hall effect can be transcribed to photonics.  Incompressibility and Landau level quantization
require fermions and charged particles, respectively, and it is not clear how one 
could introduce an effect similar to the 
Lorentz force due to a magnetic field on a system of neutral bosons.  However, a hint 
that possible analogs could exist in photonics, comes from the fact that a ``zero-field'' 
quantum Hall effect \emph{ without any net magnetic flux density} 
(and hence without Landau levels)
could occur in systems consisting of ``simple'' Bloch states with 
Broken time-reversal symmetry, as
was shown some time ago by one of us \cite{graphene}.  
The explicit ``graphene-like'' model investigated in 
Ref. \onlinecite{graphene} exploits the topological properties 
of Bloch states, which motivated us to construct 
its photonic counterpart.  This model has also turned out to be a very
useful for modeling the anomalous Hall effect in 
ferromagnetic metals \cite{nagaosa1}, and
a recently proposed ``quantum spin Hall effect''\cite{kanemele}.

While incompressibility of the fluid in the bulk quantizes the Hall conductance, 
perhaps the most remarkable feature of quantum Hall systems 
is the presence of zero energy excitations
along the \emph{edge} of a finite system.  In these edge states, electrons
travel along a single direction: this ``one-way'' propagation  is a symptom of broken time-reversal
symmetry.  In the case that the integer quantum Hall effect is exhibited by Bloch electrons,
as in the Hofstadter problem studied by TKNN\cite{tknn},
it is related to the topological  Chern invariant of the
one-particle bands and therefore.   The edge states necessarily occur at the interface between
bulk regions in which there is a gap at the Fermi energy, which have different values
of  the sum of the Chern invariants of the fully occupied bands below the Fermi level.
While the integer quantum Hall effect in such a system  itself 
involves the filling of these bands according to the
Pauli principle, and hence is essentially fermionic in nature, the existence of the 
edge states is a property of the one-electron band structure, without reference to
the Pauli principle, which suggests that this feature is 
not restricted to fermionic systems.  We have found that they indeed have a direct 
photonic counterpart, and this leads to the idea that topologically non-trivial
\emph{unidirectionally propagating photon modes} can occur along domain walls separating two
PBG regions having different Chern invariants of bands below the band gap frequency.  
In this paper, we present the formal basis of such modes, along with explicit numerical 
examples, simple model Hamiltonians, 
and semiclassical calculations confirming the concept.  

We note finally, that while Berry phases are usually associated with 
quantum mechanical interference
it can in principle occur wherever phase interference phenomena exist and are 
governed by Hermitian eigenvalue
problems, as in the case of classical electromagnetic waves in loss-free media.  

This paper is organized as follows.  In section II, we present the basic formalism of
the Maxwell normal mode problem in periodic, loss-free media,
discuss the
Berry curvature of the photonic bandstructure problem, and consider the
effects of broken time-reversal symmetry.  
In section III, we provide explicit numerical
examples of bandstructures containing non-trivial 
topological properties, and show
the occurrence of edge states along domain 
wall configurations.  Motivated by the numerical results, in 
section IV, we 
derive a simple Dirac Hamiltonian from the Maxwell equations using symmetry arguments 
as the guiding principle, and we
show how under certain conditions the \emph{zero modes} of this Dirac Hamiltonian exhibit
anomalous currents along a single direction due to the breaking of time-reversal symmetry.  
It is these zero modes that play the role of the ``gapless'' edge excitations, as we shall
consider in detail.  Section V contains semiclassical analysis, and we end with a discussion and
point out possible future directions in section VI.  
  
\section{Berry Curvature in the Maxwell Normal-mode Problem}
In this section, we discuss the formal basis of Berry curvature in the photon
band problem.  We begin with the basic formulation of the photonic 
bandstructure problem, which is somewhat more complicated than the electronic
counterpart, due to the frequency response of dielectric media.  We then 
briefly review the connection between Chern numbers, Berry curvature, and the
occurrence of gapless edge modes along the boundary where Chern numbers of a 
given band change.

\subsection{Basic Formalism}

We will be solving the source-free 
Maxwell equations for propagating electromagnetic 
wave solutions in linear, loss-free media, and will ignore absorption, 
nonlinear photon-photon interactions, and other processes which do not conserve 
photon number.  We also assume that the constitutive relations, 
reflecting the response of the media to the electromagnetic waves, are 
given by local, but spatially varying tensors with generalized frequency 
dependence.    The Berry phase, and the associated Berry curvature, are commonly
identified with quantum mechanics, but in fact are more generally associated
with the adiabatic variation of the complex eigenvectors of a Hermitian
eigenvalue system as the Hermitian matrix is varied.   

In quantum mechanics,
this Hermitian eigenvalue problem is the time-independent Schr\"odinger
equation; in the photonic context, it is the classical eigenvalue equation
for the normal modes of the Maxwell equations.    In order to make the
correspondence to the standard quantum-mechanical formulation of Berry
curvature clearer, we will use a somewhat unfamiliar Hamiltonian
formulation of Maxwell's equations, which is appropriate for loss-free
linear media.   However, we should emphasize that that our results are
in no way dependent on the use of such a formalism, and are  properties
of the Maxwell equations, however they are written.

In such a loss-free, linear medium, the coupling of electromagnetic modes 
having different frequencies can be ignored, and the electromagnetic fields 
and flux densities  $\bm X(\bm r, t)$, $\bm X \in \{\bm D, \bm B, \bm E, \bm H\}$
will be of the form
\begin{equation}
\bm X (\bm r,t) = \left( \tilde{\bm X}^*(\bm r,\omega) e^{i\omega t} +
 \tilde{\bm X} (\bm r, \omega) e^{-i \omega t}\right) ,
\end{equation}
where the quantities $\tilde{\bm X}$
are in 
general complex-valued functions of position and frequency with the 
property:
\begin{equation}
\left ( \tilde{\bm X}(\bm r,\omega)\right )^* 
= \tilde{\bm X}(\bm r,-\omega) .
\end{equation}
The dynamics 
of these fields are governed by the source-free Maxwell equations:
\begin{eqnarray}
\label{source-free}
\nabla \times \tilde{\bm{E}} = i\omega \tilde{\bm{B}} &,& 
\nabla \times \tilde{\bm{H}} = -i \omega \tilde{\bm{D}} , \label{faraday}\\
\nabla \cdot \tilde{\bm{D}} = 0 &,&
\nabla \cdot \tilde{\bm{B}} = 0 \label{gauss}.
\end{eqnarray}

Consider a single normal mode $\lambda$ propagating 
at frequency $\omega_{\lambda}$.  For the moment, ignore any internal
polarization 
or magnetization modes of the medium, and assume instantaneous, frequency-independent 
response of the dielectric material.  In this limit, the permeability 
and permittivity tensors, defined by the relations 
\begin{eqnarray}
\tilde{B}^a(\bm r,\omega_{\lambda}) = \mu^{ab}(\bm r) \tilde{H}_b(\bm r,\omega_{\lambda}) ,\\
\tilde{D}^a(\bm r,\omega_{\lambda}) = \epsilon^{ab} (\bm r) \tilde{E}_b(\bm r,\omega_{\lambda}),
\end{eqnarray}
are both positive-definite Hermitian tensors and have well-defined, positive 
definite Hermitian inverses $\epsilon^{-1}_{ab}(\bm r)$, 
$ \mu^{-1}_{ab}(\bm r)$. 
Since we have assumed a linear, loss-free medium
in which photon number is conserved, it 
is convenient to work with a Hamiltonian formalism: 
the time-averaged energy density 
of the electromagnetic radiation field is given by
\begin{equation}
\mathcal{H}^{\rm em}(\bm r) = u^e(\bm r) + u^m(\bm r) ,
\end{equation}
where
\begin{eqnarray}
\label{h0}
u^e(\bm r) = \frac{1}{2} 
\left(\tilde{\bm{D}}^*_{\lambda},\epsilon^{-1}(\bm r) \tilde{\bm{D}}_{\lambda} \right) \\
u^m(\bm r) = \frac{1}{2} 
\left(\tilde{\bm{B}}^*_{\lambda},\mu^{-1}(\bm r) \tilde{\bm{B}}_{\lambda} \right) .
\end{eqnarray}
Then, if $H^{\rm em}$ is the spatial integral of the energy density, 
the fields $\bm E$ and $\bm H$ are given by its functional
derivatives with respect to the divergence-free flux densities $\bm D$ and $\bm B$:
\begin{equation}
\delta H^{\rm em} = \int d^3{\bm r} E_a\delta B^a + H_a \delta B^a .
\end{equation}
In the local Hamiltonian formalism, the flux density fields 
$\bm D(\bm r)$ and $\bm B(\bm r)$ are the 
fundamental degrees of freedom, and they obey the following 
non-canonical
Poisson Bracket relations:
\begin{equation}
\label{pb}
\lbrace D^a(r),B^b(r') \rbrace_{\rm PB} = \epsilon^{abc} \nabla_c \delta^3 (r-r') .
\end{equation}
This Poisson bracket generates the Faraday-Maxwell equations
d(\ref{faraday}):
\begin{equation}
\frac{d\bm D}{dt} = \lbrace \bm D(\bm r), H^{\rm em}\rbrace_{\rm PB}, \quad
\frac{d\bm B}{dt} = \lbrace \bm B(\bm r), H^{\rm em}\rbrace_{\rm PB} .
\end{equation}
Note that these equations do
\emph{not} generate the Gauss law equations
(\ref{gauss}), but merely ensure that any divergences
$\nabla_aD^a$ and $\nabla _aB^a$ are
constants of the motion; the Gauss laws are additional
constraints that are compatible with the Faraday-Maxwell
equations of motion.
 
If internal polarization and magnetization modes of the
medium are ignored, a discretized form of the
electromagnetic Hamiltonian is formally identical in structure to that of 
a collection of real oscillator variables $x_i$ with non-canonical
Poisson brackets
\begin{equation}
\lbrace x_i,x_j \rbrace_{\rm PB}  = S_{ij} ,
\end{equation}
where $S_{ij}$ is a real antisymmetric matrix, and the
Hamiltonian energy function is
\begin{equation}
\label{ham0}
\mathcal{H} = \frac{1}{2} \sum_{ij}B_{ij}x_ix_j ,
\end{equation}
where $B_{ij}$ is a real-symmetric positive-definite 
matrix.   
It is useful to introduce the imaginary
antisymmetric Hermitian matrix $A_{ij}$ = $iS_{ij}$;
The canonical normal modes are given by
\begin{equation}
q_{\lambda} \pm i p_{\lambda} = (\gamma_{\lambda})^{-1}  \sum_i u^{\pm}_{i\lambda}x_i ,
\end{equation}
where $\gamma_{\lambda}$ is an arbitrary scale factor, and
where $(u_{i \lambda}^{\sigma})^*$ = $u_{i\lambda}^{-\sigma}$, $\sigma$ = $\pm$, which
obeys the generalized Hermitian eigenvalue equation
\begin{equation}
\sum_j A_{ij}u^{\pm}_{j\lambda} = \pm \omega_{\lambda} \sum_j B^{-1}_{ij} 
u^{\pm}_{j\lambda} ,
\label{discrete}
\end{equation}
with $\omega_{\lambda} > 0$, and the orthogonality condition
\begin{equation}
\sum_{ij}(u^{\sigma}_{i\lambda})^*B^{-1}_{ij} u^{\sigma'}_{j\lambda'} = 
\frac{\gamma_{\lambda}^2}{\omega_{\lambda}}
\delta_{\sigma\sigma'}\delta_{\lambda\lambda'} .
\end{equation} 
Because of the antisymmetric Hermitian property of the matrix $A_{ij}$, and
the positive-definite real-symmetric property of the matrix $B_{ij}$,
this eigenproblem  has real eigenvalues that either come in pairs,
$\pm \omega_{\lambda}$, or vanish; these equations provide a straightforward
transformation to canonical form only if the generalized eigenvalue
problem has no zero-frequency eigenvalues, which is only
the case if $A_{ij}$ is non-singular.

The coefficients $u_{i\lambda}$ are the analogs of
the electromagnetic fields
$\tilde E(\bm r, \omega)$ and $\tilde H(\bm r,\omega)$.  It is also
useful to introduce the conjugate quantities
\begin{equation}
v_{i\lambda}^{\sigma} = \sum_j B^{-1}_{ij}u^{\sigma}_{j\lambda},
\quad
\sum_i (v^{\sigma}_{i\lambda})^*u^{\sigma'}_{i\lambda'} = 
\delta_{\sigma\sigma'}\delta_{\lambda\lambda'};
\end{equation}
these are the analogs of the flux densities $\tilde {\bm D}(\bm r,\omega)$ and
$\tilde {\bm B}(\bm r,\omega)$, and $B_{ij}$ encodes the ``constitutive relations''
between analogs of fluxes and fields.

The Hamiltonian formulation of the Maxwell equations indeed presents the difficulty 
of having a null space of zero-frequency eigenvalues: by themselves, the
Faraday-Maxwell equations have static (zero-frequency) solutions
$\tilde{\bm  B } (\bm r)$  = $\bm \nabla f(\bm r)$,
$\tilde {\bm D} (\bm r )$ = $\bm \nabla g(\bm r)$; the role of
the additional Gauss law constraints is precisely  to eliminate these zero modes. 
The zero-mode problem in the Hamiltonian formulation is the counterpart of
the gauge ambiguity of the solutions of Maxwell's equations in the Lagrangian
formulation.

In the Maxwell equations,
$B_{ij}$ becomes the following positive-definite $6\times 6$ Hermitian matrix, 
\begin{equation}
B_{ij} \rightarrow  \left( \begin{array}{cc}
\epsilon^{-1}_{ab}(\bm r) & 0 \\
0 & \mu^{-1}_{ab}(\bm r)  \end{array} \right) .
\end{equation}
More precisely, this a $6\times 6$ block of an infinite-dimensional
``matrix'' that is block-diagonal in terms of the spatial coordinate $\bm r$.
(The ``$A$'' and
``$B$'' matrix  notation is common in the context of generalized Hermitian
eigenvalue problems, where the positive-definite character of the ``$B$'' matrix
guarantees reality of the eigenvalues; hopefully  the context should distinguish
our use of the symbol $B$ for such a matrix from the symbol
$\bm B(\bm r)$ used for the magnetic
flux density.)
In this continuum limit, the antisymmetric Hermitian matrix $A_{ij}$ becomes
a $6 \times 6$ matrix block  of differential operators: 
\begin{equation}
\label{a-matrix}
A^{ac}
 = \left( \begin{array}{cc}
0 & i \epsilon^{abc} \nabla_b \\
-i \epsilon^{abc} \nabla_b & 0 \end{array} \right) .
\end{equation}
This $A$-matrix can be also be elegantly expressed using the $3 \times 3$ 
spin-1 matrix representations
of the angular momentum algebra,
$\left({L}^b\right)^{ac} = i \epsilon^{abc}$:
\begin{equation}
{A} = \left(\begin{array}{cc}
0 & {\bm L}^a  \nabla_a \\
-{\bm L}^a \nabla_a & 0 \end{array}\right) .
\end{equation}

From the antisymmetry 
of A, it again follows that its eigenvalues come either in $\pm$ pairs, or are 
zero modes, corresponding to static field configurations.  Due to the 
presence of a huge band  of zero modes (one third of the spectrum), 
the $A$ matrix cannot be written in canonical form.  

Using the Poisson-Brackets, we see that the equation 
of motion of the electric and magnetic fields 
is a generalized 
Hermitian eigenvalue problem of the form 
\begin{equation*}
\left(\begin{array}{cc}
0 & {\bm L}^a  \nabla_a \\
-{\bm L}^a \nabla_a & 0 \end{array}\right)
\left [
\begin{array}{c}
\tilde{\bm E}_{\lambda} \\
\tilde{\bm H}_{\lambda}
\end{array}
\right ] 
=
\omega_{\lambda}
\left( \begin{array}{cc}
\bm \epsilon(\bm r) & 0 \\
0 & \bm \mu (\bm r)  \end{array} \right)
\left [
\begin{array}{c}
\tilde{\bm E}_{\lambda} \\
\tilde{\bm H}_{\lambda}
\end{array}
\right ]. 
\end{equation*}
In this formalism,
the energy-density
of the normal mode 
 (time-averaged over the periodic cycle)
is
\begin{equation}
u(\bm r) = 
\frac{1}{2}
\left [
\begin{array}{cc}
\tilde{\bm E}^*_{\lambda}&
\tilde{\bm H}^*_{\lambda}
\end{array}
\right ]
\left( \begin{array}{cc}
\bm \epsilon(\bm r) & 0 \\
0 & \bm \mu (\bm r)  \end{array} \right)
\left [
\begin{array}{c}
\tilde{\bm E}_{\lambda} \\
\tilde{\bm H}_{\lambda}
\end{array}
\right ] ,
\end{equation}
and the period-averaged averaged energy-current density
(Poynting vector) is
\begin{equation}
j^a(\bm r) =
\frac{1}{2}\left [
\begin{array}{cc}
\tilde{\bm E}^*_{\lambda}&
\tilde{\bm H}^*_{\lambda}
\end{array}
\right ]
\left( \begin{array}{cc}
0 & -i \bm L^a\\
i\bm L^a &  0   \end{array} \right)
\left [
\begin{array}{c}
\tilde{\bm E}_{\lambda} \\
\tilde{\bm H}_{\lambda}
\end{array}
\right ] .
\end{equation}

For practical real-space-based
calculations of the photonic normal mode
spectrum with inhomogeneous local
constitutive relations, it is very convenient
to discretize the continuum
Maxwell equations on a lattice (or network)
in a way that they
in fact have the matrix form
(\ref{discrete}), where the matrix
$A_{ij}$ reproduces the zero-mode
(null space) structure of the continuum
equations, and $H_{ij}$ represents the
local constitutive relations at network nodes
(which come in dual types, electric and magnetic).
In such a scheme, divergence-free
electric and magnetic fluxes
flow along the links of the interpenetrating
dual electric and magnetic networks, while
electromagnetic energy flows between
electric and magnetic nodes,
along the links between nodes of the network,
satisfying a local continuity equation
(see Appendix \ref{app2}).  However, there is one further
conceptual ingredient that needs to be added to 
the formalism before we can discuss
the Maxwell normal modes in ``non-reciprocal''
media which have broken time-reversal symmetry.

\subsection{Frequency dependence of the dielectric media}

The formalism discussed so far treats the constitutive
relations as static.  In general, although we will
treat them as spatially local, we cannot also treat them as instantaneous,
and must in principle treat the local permittivity and permeability
tensors as frequency-dependent, 
$\bm \epsilon (\bm r)$ $\rightarrow$ $\bm \epsilon (\bm r, \omega)$,
$\bm \mu (\bm r)$ $\rightarrow$ $\bm \mu (\bm r, \omega)$.
This is because a non-dissipative time-reversal-symmetry
breaking component of these tensors is both imaginary and antisymmetric
(as opposed to real symmetric) \emph{and} is an odd function
of frequency, making frequency-dependence inescapable in principle.

These effects can on the one hand be treated in a Hamiltonian formulation
by adding extra local harmonic oscillator modes representing local
polarization and magnetization degrees of freedom of the medium that
couple to the electromagnetic fields.     The full description of
this is again a set of harmonic oscillator degrees of freedom
described by equations of the form (\ref{discrete}).  On the
other hand, with the assumption that we are treating the electromagnetic
modes in a frequency range that is not resonant with any internal
modes of the medium ({\it i.e.}, in a frequency range where
the loss-free condition is satisfied), we can eliminate the internal
modes to yield a purely-electromagnetic description, but
with frequency dependent constitutive relations.

The details are given in Appendix \ref{app1}, but the result can be simply stated.
If all oscillator degrees of freedom are explicitly described, the eigenvalue problem
for the normal modes has the structure (\ref{discrete}), where
$B^{-1}_{ij}$ is positive-definite and real-symmetric.  This guarantees that the 
eigenvalues $\omega_{\lambda}$ are real.  However, the normal modes in some positive frequency
range $ \omega_1 < \omega < \omega_2$ can be treated by eliminating
modes with natural frequencies outside that range, to give an matrix-eigenvalue-like
problem of much smaller rank of the form
\begin{equation}
\sum_j A_{ij}u^{\pm}_{j\lambda} = \pm \omega_{\lambda} \sum_j 
B^{-1}_{ij}(\omega_{\lambda}) u^{\pm}_{j\lambda} ,
\label{selfcon}
\end{equation}
where $B_{ij}(\omega)$ is now a frequency-dependent matrix with a Kramers-Kr\"onig structure. 
The matrix $B_{ij}(\omega)$ is no longer in general real-symmetric, but
provided the eliminated modes are not resonant in the specified frequency range 
it instead is generically complex Hermitian.   
The ``eigenvalue equation'' is now a self-consistent equation:
\begin{equation}
\sum_j A_{ij}u^{\pm}_{j\lambda}(\omega) = 
\pm \omega_{\lambda}(\omega) \sum_j B^{-1}_{ij}(\omega) u^{\pm}_{j\lambda}(\omega) .
\end{equation}
This must be solved by varying $\omega$ till it coincides with an eigenvalue.
Unfortunately, while $B^{-1}_{ij}(\omega)$ is Hermitian  and non-singular
in the dissipationless frequency
range $\omega_1 < \omega < \omega_2$, it is \emph{not} necessarily positive definite,
so {\it a priori}, the eigenvalues $\omega_{\lambda}(\omega)$ are not guaranteed to
be real, except for the fact that we know that these equations were derived from a
standard frequency-independent eigenvalue problem which does have real eigenvalues.
As shown in  Appendix \ref{app1}, the Kramer-Kr\"onig structure of $B_{ij}(\omega)$ reflects
the stability of the underlying full oscillator system, giving the condition that
a modified matrix 
\begin{equation}
\tilde {B}_{ij}^{-1}(\omega)) = \frac{d}{d\omega}\left ( \omega B_{ij}^{-1}(\omega) \right )
\end{equation}
\emph{is} positive-definite Hermitian in the specified frequency range, which
is sufficient to guarantee reality of the eigenvalues in that range.  Furthermore, the quadratic
expression for the energy of a normal mode solution is given in terms of
$\tilde{B}_{ij}^{-1}(\omega_{\lambda})$ rather than $B^{-1}_{ij}(\omega_{\lambda})$,
reflecting the fact that the total energy of the mode resides in both the
explicitly-retained  degrees of freedom (the ``electromagnetic fields'')
and the those which have been ``integrated out'' (the non-resonant
polarization
and magnetization modes of the medium):
\begin{eqnarray}
x_i(t)  &=& B^{-1}_{ij}(\omega_{\lambda})u^{+}_{j\lambda}
e^{i\omega_{\lambda}t} + \text{c.c.}, \quad \omega_{\lambda} > 0, \\   
H &=& \frac{1}{2}\sum_{ij}\tilde{B}^{-1}_{ij}(\omega_{\lambda})(u^{+}_{i\lambda})^*u^{+}_{j\lambda},
\quad \frac{d H}{dt} = 0.
\end{eqnarray}

If the frequency dependence of the positive-definite Hermitian matrix $\tilde {B}_{ij}(\omega)$ 
is negligible in the range $\omega_1 < \omega < \omega_2$, so $\tilde {B}_{ij}(\omega)$ $ \approx$
$\tilde {B}_{ij}(\omega_0)$, with $\omega_0$ = $(\omega_1+\omega_2)/2 $, one can replace
$B^{-1}_{ij}(\omega)$ in (\ref{selfcon}) by the positive-definite frequency-independent
Hermitian matrix $\tilde B^{-1}_{ij}(\omega_0)$.
This in turn allows the eigenvalue problem to be transformed into the 
standard Hermitian eigenvalue problem
\begin{equation}
H_{ij}
w_{j}^{(\lambda)} = \omega_{\lambda}w_{i}^{(\lambda)},
\quad \left ( \bm w^{(\lambda )},\bm w^{(\lambda')}\right )
 = \delta_{\lambda\lambda'} ,
\end{equation}
with scalar product
\begin{equation}
\left (\bm w,\bm w'\right )
\equiv \sum_j w^*_{j} w_{j}',
\end{equation}
valid for positive $\omega_{\lambda}$ in the frequency range where 
$\tilde {B}_{ij}(\omega)$ $ \approx$$ \tilde {B}_{ij}(\omega_0)$, with
\begin{equation}
H_{ij} = \left ( \tilde{B}^{1/2}(\omega_0)A\tilde{B}^{1/2}(\omega_0) \right )_{ij},
\end{equation}
and
\begin{equation}
\label{defw}
u^+_{i\lambda} \propto \sum_j \tilde B^{1/2}_{ij}(\omega_0) w_{j\lambda}.
\end{equation}
This allows well-known Berry-curvature formulas from the standard Hermitian
eigenproblem \cite{simon} to be quickly translated into the generalized problem.
It turns out that when the full problem with frequency-dependent
constitutive relations is treated, the standard formula for the
\emph{Berry connection} remains correct
with the simple replacement $\tilde {B}_{ij}(\omega_0)$  
$\rightarrow$ $\tilde {B}_{ij}(\omega_{\lambda})$ ( the Berry curvature
and Berry phase can both be expressed in terms of this Berry
connection).

\subsection{Berry curvature in Hermitian eigenproblems}

Let $H_{ij}(\bm g )$ be a family of complex Hermitian matrices
defined on a manifold parameterized by  a set $\bm g$ of independent coordinates
$g^{\mu}$, $\mu$ = $1,\ldots D$    It is assumed that the matrix is generic,
so its eigenvalues are all distinct; as it is well known, three independent parameters
must be ``fine-tuned'' to produce a ``accidental degeneracy'' between a pair of eigenvalues.
Thus if the parametric variation of the Hermitian matrix 
is confined to a  two-parameter submanifold,
each eigenvalue $\omega_{\lambda}(\bm g)$ will generically remain distinct.  
Under these circumstances, the
corresponding eigenvector is fully defined by the eigenvalue equation and normalization 
condition, up
to multiplication by a unimodular phase factor, that can vary on the manifold:
\begin{equation}
w_{\lambda i}(\bm g )\rightarrow  e^{ i \phi (\bm g )} w_{\lambda i}(\bm g).
\end{equation}
This is the well-known ``$U(1)$ gauge ambiguity'' of the complex Hermitian eigenproblem.
Associated with each eigenvector is a gauge-field (vector potential in the 
parameter space), called  the ``Berry connection'': 
\begin{equation}
\mathcal A_{\mu}^{(\lambda )}( \bm g) = 
-i\left (\bf w_{\lambda}(\bm g)),\partial_{\mu}\bm w_{\lambda}(\bm g) \right ),
\quad \partial_{\mu} \equiv \frac{\partial}{\partial g^{\mu}}.
\end{equation}

This field on the manifold is gauge-dependent, like the electromagnetic vector 
potential $\bm A(\bm r)$,
but its curvature (the ``Berry curvature''), the analog of the magnetic flux density 
$\bm B(\bm r)$ = $\bm \nabla \times \bm A(\bm r)$,
is gauge invariant and given by
\begin{equation}
\mathcal F_{\mu\nu}^{(\lambda)}(\bm g) =  \partial_{\mu}\mathcal A^{(\lambda)}_{\nu}(\bm g) 
- \partial_{\nu}\mathcal A^{(\lambda)}_{\mu}(\bm g).
\end{equation}
The Berry phase associated with a closed path $\Gamma$  in parameter space
is given (modulo $2\pi$) by
\begin{equation}
\exp\left(-i \phi^{(\lambda)}(\Gamma) \right) = \exp\left(-i \oint_{\Gamma} 
\mathcal A_{\mu}(\bm g)dg^{\mu} \right).
\end{equation}.

Ignoring frequency-dependence, the oscillator system has 
\begin{equation}
\bm H(\bm g) =   {\bm 
B}^{1/2}(\bm g) \bm A  {\bm  B}^{1/2} (\bm g),
\end{equation}
where the positive-definite Hermitian
matrix ${\bm B}(\bm g)$ 
can continuously vary as a function of some parameters
$\bm g$, but $\bm A$ is invariant.   Then converting to
the oscillator variables gives 
\begin{eqnarray}
\label{generalconnection}
\mathcal A^{(\lambda )}_{\mu}(\bm g) 
&=&  {\rm  Im.}
\left ( 
\frac 
{\left ( \bm u^{(\lambda )}
,\tilde{\bm B}^{-1}(\bm g,\omega_{\lambda}) 
\partial_{\mu} 
\bm u^{(\lambda)}\right )}
{\left ({\bm u}^{(\lambda )}
,\tilde{\bm B}^{-1}(\bm g,\omega_{\lambda})  \bm u^{(\lambda)}\right )}
\right ).
\end{eqnarray}
Here $\bm B^{-1}(\bm g)$ has been replaced by
$\tilde{\bm B}^{-1}(\bm g,\omega_{\lambda})$ which is the correct
result when frequency dependence of $\bm B^{-1}(\bm g,\omega)$ is
taken into account (see Appendix \ref{app1}).

\subsection{Photonic bands and Berry curvature}

In the case of periodic systems, the normal modes have discrete translational
symmetry classified by a Bloch vector $\bm k$ defined in the Brillouin zone, 
\textit{i.e.}, defined modulo a reciprocal vector $\bm G$.
For fixed $\bm k$, the spectrum of normal mode frequencies $\omega_n(\bm k)$ is
discrete, labeled by band indices $n$,  
and, as emphasized by Sundaram and Niu\cite{sundaram niu} in the electronic
context, the Bloch vector of a wavepacket plays the role of the control-parameter
vector $\bm g$.

In order to compute the Berry curvature of the photon band
Bloch states, we shall find it 
convenient to work in a fixed Hilbert space for all Bloch vectors 
$\bm{k}$, and we do this by
performing a unitary transformation on the $\bm{A}$
``matrix'' (which becomes the $6\times 6$ matrix of differential operators 
(\ref{a-matrix}) in the continuum
formulation of the Maxwell equations) as
\begin{equation}
{\bm{A}}(\bm k,\bm \nabla) \equiv
  e^{-i \bm{k} \cdot \bm{r}} \bm{A(\bm \nabla)} 
e^{i\bm{k} \cdot \bm{r}}  = \bm A(\bm \nabla  + i\bm k).
\end{equation}
Note that parametric dependence on the Bloch vector $\bm k$ is a little different from
parametric dependence on parameters $\bm g$ that control the Hamiltonian.
After projection into a subspace of fixed $\bm k$, the
``$\bm A$'' matrix also becomes parameter-dependent, while (if the constitutive
relations are taken to be completely local) the ``$\bm B$''
matrix in the photonics case is 
only implicitly $\bm k$-dependent through its self-consistent dependence
on the frequency eigenvalue.
(Parameter-dependence of the ``$\bm A$'' matrix  
does not affect the expression (\ref{generalconnection}) for the
Berry connection.)

The discrete eigenvalue spectrum $\omega_n(\bm k)$ 
is then obtained by solving the self-consistent 
matrix-differential-equation eigenvalue
problem:
\begin{equation}
\bm A(\bm k,\bm \nabla) \bm u_{n}(\bm{k},\bm r)  = \omega_n(\bm k) 
\bm B^{-1}(\bm r ,\omega_n(\bm k))\bm u_{n}(\bm{k},\bm r) ,
\end{equation}
where $\bm B^{-1}(\bm r,\omega)$ is the $6 \times 6$ block-diagonal
permittivity-permeablity tensor
${\rm diag}(\bm \epsilon(\bm r, \omega),\bm \mu (\bm r,\omega))$,
and  $\bm u_n(\bm k,\bm r)\exp i \bm k\cdot \bm r$ 
represents the 6-component complex vector
$(\tilde {\bm E}_n(\bm k,\bm r), \tilde {\bm H}_n(\bm k,\bm r))$ of
electromagnetic fields of the normal mode with Bloch vector $\bm k$
and frequency $\omega_n(\bm k)$.
The eigenfunction satisfies the periodic boundary condition
$\bm u_n(\bm k,\bm r  + \bm R)$ = $\bm u_n(\bm k,\bm r)$, where
$\bm R$ is any lattice vector of the photonic crystal, where
$\bm B^{-1}(\bm r+\bm R, \omega)$ = $\bm B^{-1}(\bm r, \omega)$.

The transcription of Eq.(\ref{generalconnection}) 
to the case of periodic media then gives the three-component Berry connection
(vector potential) in $\bm k$-space as 
\begin{equation}
\mathcal{A}^a_{(n)}(\bm{k}) = {\rm Im.} \left( 
\frac{\left( \bm{u}_n(\bm k), \tilde{\bm{B}}^{-1}(\omega_n(\bm k)) \nabla_k^a \bm{u}_n(\bm k) \right) }
{\left( \bm{u}_n(\bm k), \tilde{\bm{B}}^{-1}(\omega_n(\bm k)) \bm{u}_n(\bm k) \right)}
\right ) .
\label{photonberry}
\end{equation}
The scalar products in Eq.(\ref{photonberry}) are defined by the
trace over the six components of $\bm u_n(\bm k,\bm r)$, plus integration
of the spatial coordinate $\bm r$ over a unit cell of the photonic
crystal.   By construction, if a ``Berry gauge transformation''
$\bm u_n(\bm k,\bm r)$ $\rightarrow$ $\bm u_n(\bm k,\bm r)\exp i\chi_n(\bm k)$
is made, $\mathcal A^a_{(n)}(\bm k)$ $\rightarrow$ $\mathcal A^a_{(n)}(\bm k)$
+ $\nabla^a_k\chi_n(\bm k)$.
 
In three-dimensional $\bm k$-space, the antisymmetric
Berry curvature tensor $\mathcal F^{ab}_{(n)}(\bm k)$
= $\nabla_k^a\mathcal A^b_{(n)}(\bm  k) -\nabla_k^b\mathcal A^a_{(n)}(\bm  k)$
can also be represented as the three-component
``Berry flux density'' $\Omega^{(n)}_a(\bm  k)$  = 
$\epsilon_{abc}\nabla_k^b\mathcal A^c_{(n)}(\bm k)$
(the  $\bm k$-space curl 
of the Berry connection), to emphasize the duality between
$\bm r$-space and $\bm k$-space, and the analogy between Berry flux
in $\bm k$-space and magnetic flux in $\bm r$-space,

If a wavepacket travels adiabatically (without
interband transitions) through a region with slow
spatial variation of the properties of the medium, so the photonic normal-mode 
eigenvalue
spectrum can be represented as a position-dependent dispersion
relation $\omega_n(\bm k,\bm r)$, the wavepacket must be
accelerated as its mean Bloch vector $\bm k$ slowly changes to keep its frequency
constant.
When translated into the language of photonics,
the semiclassical electronic equations of motion then become the 
equations of ray
optics:
\begin{eqnarray}
\hat n^a\frac{dk_a}{dt} &=& -\hat n^a \nabla_a\omega_n(\bm k,\bm r),
\\
\frac{dr^a}{dt} &=& \nabla_k^a\omega_n(\bm k,\bm r) +
\mathcal F^{ab}_n(\bm k,\bm r)\frac{dk_b}{dt},
\label{anomalous}
\end{eqnarray}
where $\hat n$ $\propto $ $d\bm r/dt$ is parallel to the ray path, 
$\nabla_a$ $\equiv$ ${\partial}/{\partial r^a}$
and $\nabla^a_k$ $\equiv$ ${\partial}/{\partial k_a}$ (it is useful
to use covariant and contravariant indices to distinguish components of
spatial coordinates $r^a$ from the dual Bloch vector  components $k_a$).
The Bloch-space Berry curvature $\mathcal F^{ab}_n(\bm k,\bm r)$ controls the
additional ``anomalous velocity\cite{karplusluttinger}'' correction
in (\ref{anomalous})
to the familiar group velocity of a
wave packet $v_n^a(\bm k)$ = $\nabla_k^a\omega_n(\bm k)$, which is
active only when the wavepacket is being accelerated by the
inhomogeneity of the medium.

Before we conclude our general discussion on Berry curvature in photon band
systems, we must state the constraints inversion and time-reversal symmetries place 
on the tensor $\mathcal{F}^{ab}_n(\bm k)$.  In what follows, we will use the Bloch 
state $\bm w_n$ defined in Eq.(\ref{defw}).    
If inversion symmetry (I) is present, the periodic part of the Bloch state 
$\bm w_n(\bm{k}$ has the following property: $\bm w_n(\bm{k}) = \bm w_n(-\bm{k})$
whereas if time-reversal symmetry (T) is present, $\bm w_n(\bm{k}) = \bm w_n^*(-\bm{k})$.  
If only (I) is present, it then follows that 
$\mathcal{F}_n^{ab}(\bm{k}) = \mathcal{F}_n^{ab}(-\bm{k})$, whereas if only (T) is present, 
$\mathcal{F}_n^{ab} = - \mathcal{F}_n^{ab}(-\bm{k})$.  If \textit{both} 
symmetries are present, then the Berry Curvature is identically zero 
everywhere except at isolated points of ``accidental degeneracy'', where 
it is not well-defined.  When $\mathcal{F}_n^{ab}$ is non-zero, the 
phases of the Bloch vectors cannot all be chosen to be real.  These 
properties will be crucial when we consider the effects of 
various symmetry breaking perturbations on the photon band structure.

\subsection{Topological structure of the photon bands}

The main consequence of having bands with non-zero Berry curvature field 
is that if the path C is closed and encloses an entire 
Brillouin zone, the single-valuedness of the 
state $\bm w_n$ requires that 
\[ \exp \left( \oint \mathcal{A}_n^a dk_a \right) = 
\exp \left( \int \int dk_a \wedge dk_b \mathcal{F}_n^{ab} \right) = 1 \]
or, 

\begin{equation}
\int \int_{S_C} dk_a \wedge dk_b \mathcal{F}^{ab}_n = 2 \pi \mathcal{C}^{(1)}_n ,
\end{equation}
where $\mathcal{C}^{(1)}_n$ is an integer, known as the \textit{Chern invariant} 
associated with the nth band, and have 
well-known consequences in the quantum Hall effect: in the integer quantum 
Hall effect, where the interactions among electrons are weak, the Hall 
conductance is expressed in terms of the sum of all Chern invariants of bands 
below the Fermi level \cite{tknn}:

\begin{equation}
\sigma_{H} = \frac{e^2}{2 \pi \hbar} \sum_{i, \epsilon_i < \epsilon_f} 
\mathcal{C}^{(1)}_i .
\end{equation}

The gauge structure of the photon band problem outlined above
 is formally analogous
to the local $U(1)$ gauge invariance of ordinary electromagnetism.  
Note that the gauge invariance refers to the phase of the of the six-component
electromagnetic fields as a whole; adding arbitrary phase 
$\bm{k}$-dependent
phase factors to each field separately will in general not preserve the 
Maxwell equation constraint.  

A phase convention can be specified, for instance, by arbitrarily 
choosing real and imaginary axes of the phases; the local gauge-dependent 
phase fields of the electromagnetic Bloch states are then represented as  
two-component rotor variables at each point of the Brillouin zone.  In 
addition, a gauge choice may be made separately for each band so long 
as the spectrum remains non-degenerate.  

By representing the phase covering on the Bloch manifold this way, 
the possibility of the occurrence of topological
defects of the gauge field become transparent.  Local gauge transformations 
correspond to local smooth deformations of the rotor variables, and the Chern 
invariant corresponds to the total winding number of theses rotor variables 
along a closed path enclosing the entire Brillouin zone.  

In the case of two dimensional
Bloch bands, the defects of the phase field 
are point singularities, having a zero-dimensional ``core'' region where 
a phase convention is not well defined, due to quasi-degeneracies with 
neighboring bands.  It is clear that Bands can have non-zero Chern numbers 
only if time-reversal symmetry is broken.  Otherwise, the Berry curvature 
will be an odd function of $k$, and it's integral over the entire 2D Brillouin 
zone vanishes.

In three dimensions, the defects of the phase 
field are line defects or ``vortices'' and their stability requires quasi-
degeneracies to occur along isolated \textit{lines} in reciprocal space.  

\begin{figure}[htbp]
\begin{center}
\includegraphics[width=3.4in,keepaspectratio,clip]{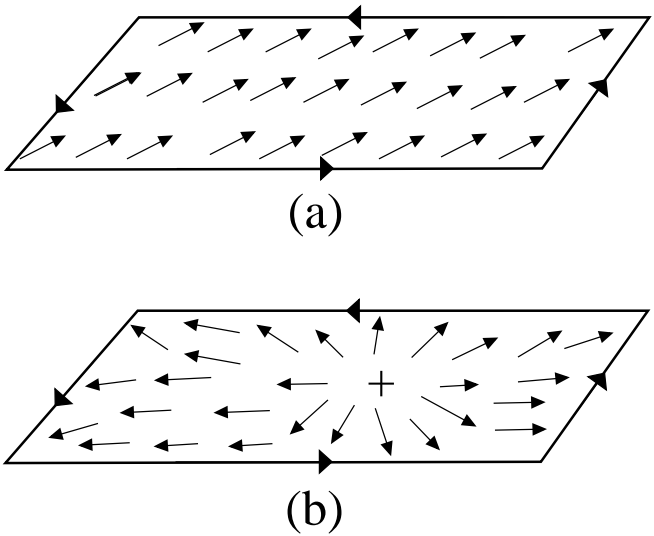}
\end{center}
\caption{A representation of the phase fields of the photonic Bloch states 
in a two dimensional Brillouin zone using two component rotors.  The 
entire set of six electric and magnetic fields is associated with a single 
phase at each point in the Brillouin zone.  The Chern invariant simply 
represents the winding number of this phase along the Brillouin zone boundary 
and is also given by the integral of the Berry curvature $\mathcal{F}^{xy}$ 
over the two dimensional Brillouin zone.  The phases in (a) correspond 
to bands with both inversion and time-reversal symmetries, and  
the phases of the band can be chosen to be real every where in the Brillouin 
zone.  For bands having non-zero Chern invariants (b), the phase 
around the zone boundary winds by an integer multiple of $2\pi$ and there is 
a phase vortex-like  
singularity somewhere in the Brillouin zone where the Berry connection 
cannot be defined, due to the occurrence of quasi-degeneracies.}
\end{figure}

In the photonic system of interest, even if photon bands can have non-zero 
Chern numbers, there can be no Hall conductance as given above due to their 
Bose statistics (and hence, to their finite compressibility).  However, the 
connection between \textit{edge modes} and Chern invariants is independent of 
statistics: if the Chern number of a band changes at an interface, the 
net number of unidirectionally moving modes localized at the interface is 
given by the \textit{difference} of the Chern numbers of the band at the 
interface.  We shall consider the problem of how 
Chern numbers can change across an interface in the next section.  

\begin{figure}[htbp]
\begin{center}
\includegraphics[width=3.4in,keepaspectratio,clip]{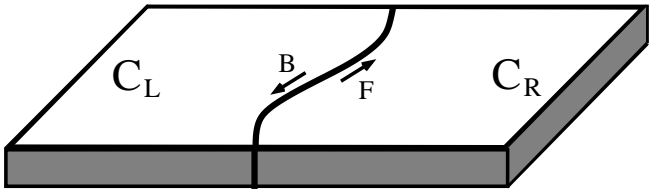}
\end{center}
\caption{The number of forward minus the number of backward moving edgemodes
equals the difference of the Chern number of the band across the interface.}
\end{figure}

Since the Chern invariant of a band is a topological number, 
it therefore cannot vary smoothly as we vary some parameter of the 
periodic eigenproblem.  So long as a band remains non-degenerate, it's 
Chern number cannot vary.  However, if we tune some parameter $\lambda$ of the 
Hamiltonian to a critical value $\lambda_c$ such that two bands 
having non-zero Chern invariants touch 
at some isolated point in the Brillouin zone when $\lambda = \lambda_c$, 
the two bands can exchange their Chern numbers 
at these degenerate points; if we tuned $\lambda$ beyond its critical value, 
the bands would emerge with different Chern invariants.  Since the total 
Berry ``magnetic flux'' of all bands remains fixed always, if only two 
isolated bands exchange their Chern numbers at points of degeneracy, 
the sum of their Chern numbers must remain invariant \cite{simon}.  

Generically, 2D bands with both time-reversal and inversion symmetry 
touch at isolated points of accidental degeneracy 
in a linear conical fashion, 
forming ``Dirac cones'' in the vicinity of which the
 spectrum is determined by a massless Dirac Hamiltonian:

\begin{equation}
 \mathcal{H} \equiv \omega - \omega_D = v_D \left( \delta k^1 \sigma_1 + 
\delta k^2 \sigma_2 \right) ,
\end{equation}
where $v_D$, is a parameter that gives the slope of the cone close to the 
accidental degeneracy.  

\begin{figure}[htbp]
\begin{center}
\includegraphics[width=3.4in,keepaspectratio,clip]{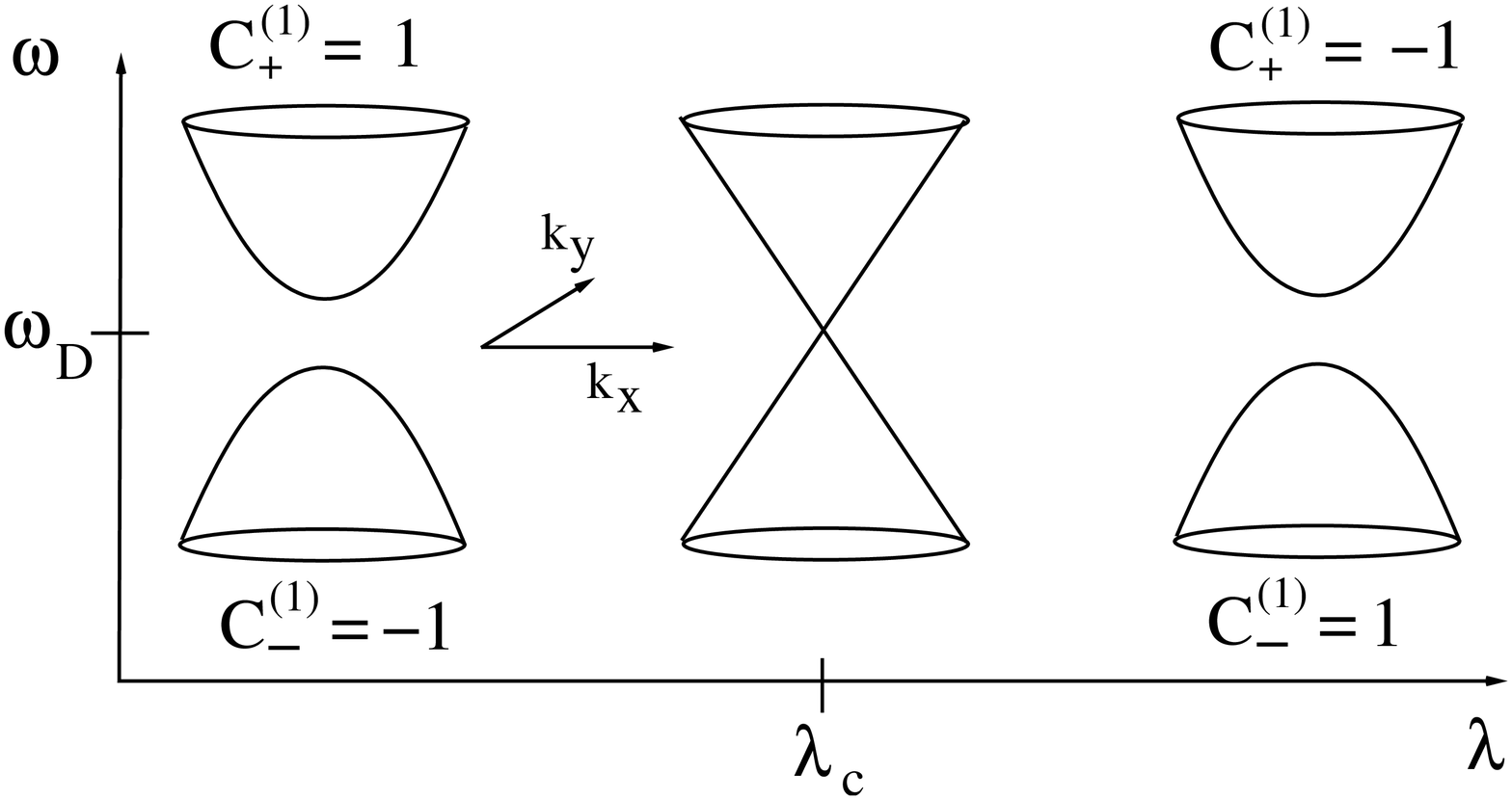}
\end{center}
\caption{As we tune some parameter $\lambda$ of the Hamiltonian across a 
critical point where ``accidental degeneracies'' occur, and two bands touch 
in a linear fashion forming a ``Dirac point'', Chern numbers of bands may 
be exchanged.}
\end{figure}

\section{Broken T and I in photonics}

In this section, we shall discuss our strategy for constructing photon
bands with non-zero Chern invariants, and ``chiral'' edge states, 
whose existence is confirmed in the following sections.  

To break time-reversal symmetry in photonics, we shall need magneto-optic 
materials (i.e. a Faraday rotation effect).  Such materials are characterized
by their ability to rotate the plane of polarization of light, when placed
in a magnetic field, and are used in
conventional optical isolators.  The amount of rotation per length 
is known as a 
Verdet coefficient, which depends on temperature as well as 
on the wavelength of 
light.  Materials known to have large Verdet 
coefficients ($\sim 10^0 mm^{-1}$ at 
wavelengths of the order of microns) are the iron garnet 
crystals such as $\mathrm{Ho_3 Fe_5 0_{12}}$
\cite{faraday}.  Due to the breaking of time-reversal 
symmetry in this materials, 
the eigenfrequency degeneracy is lifted for light 
characterized by different states 
of circular polarization.  

\begin{figure}[htbp]
\begin{center}
\includegraphics[width=3.4in,keepaspectratio,clip]{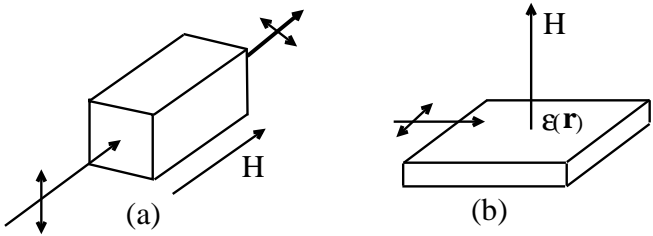}
\end{center}
\caption{In the conventional Faraday effect used in optical isolators, light travels
in the same direction as the applied field, resulting in the rotation of its 
polarization plane.  However, in the photonic analog of a 2DEG heterojunction,
light travels in F  }
\label{polarizer}
\end{figure}

While such magneto-optic devices employ magnetic fields in the direction of travel of the light 
beam, we shall be interested in two dimensional photonic crystals with the magnetic fields
placed perpendicular to the plane of propagation of light, as shown in Fig. \ref{polarizer}.  
We shall call the axis perpendicular to the 2D photon bands the Faraday axis,
and the setup here is reminiscent of a 2D electron gas placed in a perpendicular
magnetic field.  

Although we now have a means of introducing broken time reversal symmetry, we
still need a strategy for the nucleation of equal and opposite pairs of 
Chern invariants on bands near points of accidental degeneracy.  
To do this, we choose 
hexagonal lattice geometry.  The threefold rotation symmetry of such a 
system guarantees the existence of Dirac points in the Brillouin zone corners
when both inversion and time-reversal symmetry are \textit{present}; in this case the only 
irreducible representations of the space group of three-fold rotations 
correspond to non-degenerate singlets and degenerate doublets.  As a simple
example consider the case of free photon ``bands'' with dispersion 
$\omega=c| \bm{k}|$ in the first Brillouin zone of a triangular lattice.  The
eigenfrequencies of the photons are six-fold degenerate at the zone corners.
Adding a weak periodic perturbation in the constitutive relations will
lift the degeneracy and the bands will now be either non-degenerate or will
form degenerate doublets, as demanded by the symmetry of the perturbation.  
Due to the 6-fold rotation symmetry, the doublets are allowed to have a 
linear dispersion close to the
zone corners and shall be our Dirac points of interest, whereas the 
non-degenerate singlet bands disperse quadratically.  
We shall provide explicit examples of hexagonal photonic bandstructures having 
Dirac points in section V.  

While the existence of such Dirac points are virtually guaranteed in 
triangular lattice systems, their \textit{stability} has little to do with
lattice geometry.  Such points are stable in two dimensions only because of
the presence of time-reversal symmetry and inversion symmetry, when the 
eigenvalue problem is a real-symmetric one: in this case it is possible 
to find ``accidental'' degeneracies by varying just two parameters, according 
to the Wigner-von Neumann Theorem.  
Thus, if the perfect hexagonal geometry of the constitutive relations 
is slightly distorted, the Dirac points will simply move elsewhere in the 
two dimensional Brillouin zone.  Provided that such distortions are not
too strong such that an axis of two-fold rotations is introduced, in which
case the linear dispersion characteristic of a Dirac point is no longer 
allowed,
or if the distortion is so great that the Dirac points meet and  
annihilate at a point of inversion symmetry, Dirac points will still exist
in the system.

If, however, inversion or time-reversal symmetry is broken in the system, 
the eigenvalue problem becomes complex Hermitian, and according to the 
Wigner-von Neumann theorem, three parameters are required
to ensure stability of the Dirac points - in this case, the Dirac point degeneracy 
of the 2D bandstructure is immediately lifted.  In both cases, the two bands 
which split apart each
acquire a non-zero Berry curvature field $\mathcal{F}_{xy}(\bm{k})$.  

If inversion symmetry alone is broken, $\mathcal{F}_{xy}(\bm{k})$ is an
\textit{odd} function of $\bm{k}$, as discussed above.  While the bands
do have interesting semiclassical dynamics due to their anomalous velocity,
they do not have any interesting topological properties since their 
Chern invariants are identically zero.  

On the other hand, if time-reversal symmetry alone is broken, via the Faraday
coupling, the Berry curvature field will be an $\textit{even}$ function of
$\bm{k}$, and each band which split apart due to the Faraday coupling
will have equal and opposite non-zero Chern invariants.  

Finally, if we can slowly tune the Faraday coupling in \textit{space}, 
from a positive value,
across the critical value of zero, where the local bandstructure problem
would permit Dirac spectra, to a negative value, we would generate a system
of photonic bands with non-zero Chern numbers, that get exchanged at 
the region of space corresponding to the critical zero Faraday coupling.  
It then follows, that modes with exact correspondence to the integer quantum
Hall edge states would arise in such a system.  In the following section, we
shall display this explicitly using an example Bandstructure.

\section{Explicit Realization of Edge modes}
\begin{figure}[htbp]
\begin{center}
\includegraphics[width=3.4in,keepaspectratio,clip]{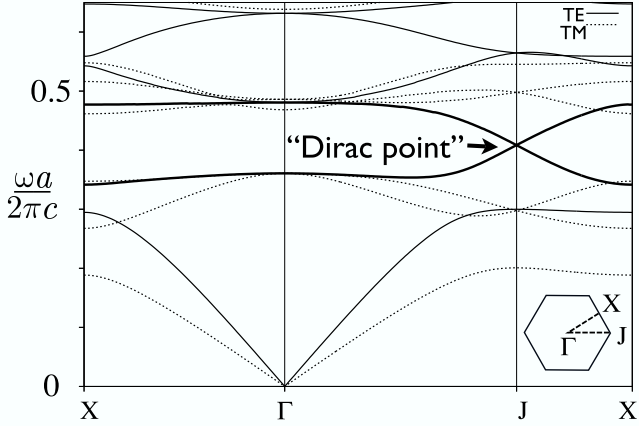}
\end{center}
\caption{Photon bands in the $k_z=0$ plane of a 2D hexagonal lattice 
of cylindrical dielectric rods.  Electromagnetic waves are propagating only 
in the $x-y$ plane (Brillouin zone shown in the lower right).  As in Ref. 
\onlinecite{maradudin}, the rods have a filling fraction $f=0.431$, $\epsilon=14$,
 and the background has $\epsilon=1$.  The band structure contains a pair of 
Dirac points at the zone corners (J).}
\label{band}
\end{figure}

An example of a photonic bandstructure with the desired properties is shown in
Fig. \ref{band}.   It consists of a triangular lattice of dielectric rods 
($\epsilon_a = 14$) placed in a background of air ($\epsilon = 1$) 
with a area filling ratio of $f=0.431$.  The authors of
ref. \onlinecite{maradudin}, in a quest for optimal 
photonic bandgap materials, first studied this system.  
They computed the TE mode spectrum and 
found a full band gap in the TE spectrum.  We have reproduced 
numerically their calculation and have also computed the spectrum 
for the TM modes.

The key feature of this particular system which is of importance
to us are the presence of a pair of Dirac points in the spectrum
of the TE modes which are well isolated from both the remaining TE 
and TM modes.  Each of the six zone corners contains the Dirac cone 
spectrum, but there are only two distinct
Dirac points, the others being related by reciprocal 
lattice translations of these points.  In this
particular system, the two Dirac points are related by inversion in k-space.

As we have discussed, a gap immediately opens when either inversion or 
time-reversal symmetries are broken in this system.  We break inversion 
symmetry in the simplest possible way by introducing a slight imbalance in
the value of the dielectric tensor in the rods at opposite ends of the unit 
cell, and we parameterize the inversion breaking by defining the quantity

\begin{equation}
M_I = \log \left(\frac{\epsilon_+}{\epsilon_-} \right),
\end{equation}
where $\epsilon_+ (\epsilon_-)$ is the value of the permittivity inside the
rods in the upper (lower) half of the unit cell depicted in Fig. \ref{band}.

To break time-reversal symmetry, we add a faraday effect 
term in the region outside the rods.  This is done by 
giving the dielectric tensor a slight 
imaginary component without varying the constitutive relations 
inside the rods:

\begin{equation}
\mbox{outside rods: }\epsilon_{ij}^{-1}(\bf{x}) = \left( \begin{array}{cc}
\epsilon_b^{-1} & i\Lambda \\
-i\Lambda & \epsilon_b^{-1} \end{array} \right) ,
\end{equation}

\begin{equation}
\mbox{inside rods: }\epsilon_{ij}^{-1}(\bf{x}) = \left( \begin{array}{cc}
\epsilon_a^{-1} & 0 \\
0 & \epsilon_a^{-1} \end{array} \right).
\end{equation}

We also define a parameter 

\begin{equation}
M_T = \Lambda ,
\end{equation}
to define the strength of the time-reversal symmetry breaking perturbation.  

\begin{figure}[htbp]
\begin{center}
\includegraphics[width=3.4in,keepaspectratio,clip]{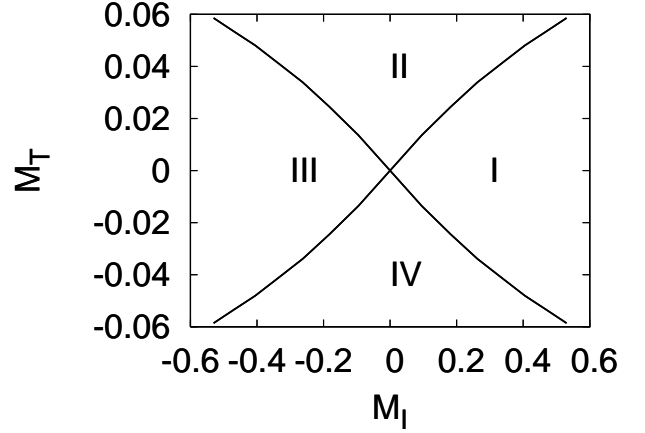}
\end{center}
\caption{Phase diagram of the photonic system as a function of inversion and time-reversal
symmetry breaking.  In regions I and III, the gap openings of both Dirac points are 
primarily due to inversion symmetry breaking, whereas in regions II and IV, the breaking of
time-reversal symmetry lifts the degeneracy of the bands which formed the Dirac point.  In 
all four regions, the two bands of interest have non-zero Berry curvature, but only in 
regions II and IV do they contain non-zero Chern numbers.}
\label{pd}
\end{figure}

We first determined the phase diagram of the system in the $(m_I,m_T)$ plane by breaking
both inversion \textit{and} time-reversal symmetry, and locating 
special values of the symmetry breaking parameters that result in the 
closing of the bandgap at one or more Dirac points (Fig. \ref{pd}).
The phase diagram separates regions characterized by bands just below the 
band gap having a non-zero Chern number from regions with all bands having
zero Chern numbers.  The boundary between these phases are where the gap
at one or more of the Dirac points vanishes, as shown in Fig. \ref{pd}.   
Since there are two Dirac points, each phase boundary corresponds to 
the locus of parameters for which  the gap at one of the Dirac points closes.
Thus, both Dirac points close only when both lines intersect, namely at the
point $(m_I=0,m_T=0)$, where both inversion and time-reversal symmetries 
are simultaneously present.  
When inversion symmetry alone is broken, the Berry curvature 
field of Dirac point 1 is 
equal in magnitude and opposite in sign of the Berry 
curvature at the second Dirac point.  When
time-reversal symmetry is broken, on the other hand, 
each Dirac point has an identical (both in
magnitude and sign) Berry curvature field.  
In this case, the photon bands which split apart at 
the Dirac point each have Non-zero Chern number, 
which depends only on the direction of the Faraday
axis ($\pm \hat{z}$).  

\begin{figure}[htbp]
\begin{center}
\includegraphics[width=3.4in,keepaspectratio,clip]{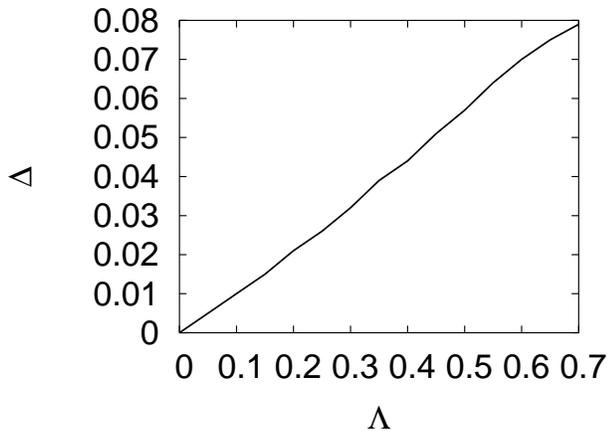}
\end{center}
\caption{The bandgap opened by time-reversal breaking as a function of the strength 
of the Faraday coupling shows that the gap is linearly proportional to $\Lambda$.}
\label{gap}
\end{figure}

We have also studied numerically the frequency gap as a function 
of the time-reversal breaking perturbation above and found that so long as the
dielectric tensor remains positive-definite, the gap increases linearly 
with $\epsilon_{xy}$ (\ref{gap}).  This will be important when we 
consider effective Dirac 
Hamiltonians for this problem: as we shall see, 
the fact that exactly at the zone corner, the gap rises linearly with $M_T$ is 
consistent with the spectrum of a massive Dirac Hamiltonian with mass $M_T$.  
Thus, we have shown an example of a bandstructure which contains Dirac points 
whose gaps can be tuned
using time-reversal and inversion symmetry breaking perturbations.  
We can now show the existence 
of ``chiral'' edge states in this system.  

To study edge states in this system, we introduce a ``domain wall'' configuration across which 
the Faraday axis reverses.  As we shall now show numerically, (and justify analytically 
in the following section), the edge modes that occur along the domain wall are bound states
that decay exponentially away from the wall while propagating freely in the direction parallel 
to the interface.  In order to study the exponential decay of these modes, we glue together 
$N$ copies of a single hexagonal unit cell along a single lattice translation direction 
$\bm{R}_{\bot}$, which shall be the direction perpendicular to the domain wall.  We treat this
composite cell as a unit cell with periodic boundary conditions.  Since a domain corresponds to
a certain direction of the Faraday axis, we study a configuration here in which the axis 
changes direction abruptly across the domain wall from the $+ \hat{z}$ to the $-\hat{z}$ direction.  

\begin{figure}[htbp]
\begin{center}
\includegraphics[width=3.0in,keepaspectratio,clip]{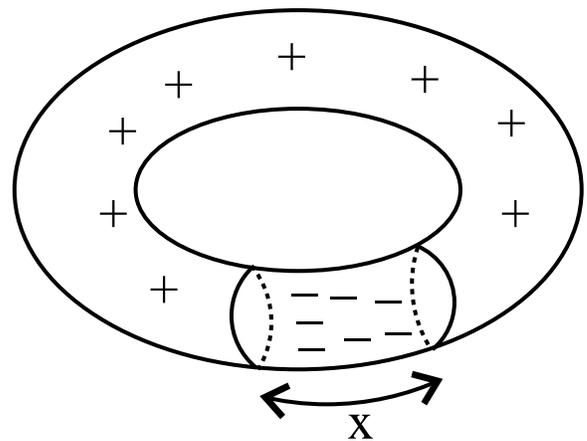}
\end{center}
\caption{In a system with periodic boundary conditions, there are necessarily
two domain walls separating regions with different faraday axes.  We study
the gap of the spectrum at the (now non-degenerate) Dirac points as a
function of the distance $\bm x$ between the two walls.}
\label{torus}
\end{figure}

\begin{figure}[htbp]
\begin{center}
\includegraphics[width=3.0in,keepaspectratio,clip]{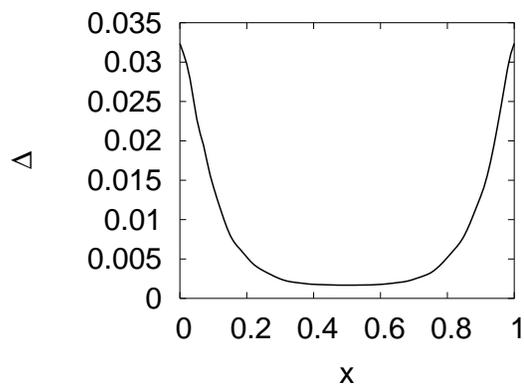}
\end{center}
\caption{The spectral gap between the two bands which split apart due to the 
breaking of time-reversal symmetry.  The spectrum is computed on a torus 
for the extended system consisting of 30 copies of the hexagonal unit cell.  
Furthermore, domain walls, across which the sign of the Faraday axis flips 
are introduced, and the spectrum is plotted as a function of the separation 
$\bm x$ between the walls (see also Fig. \ref{torus}).}    
\label{gapofx}
\end{figure}

When we consider the spectrum on a torus, 
there are necessarily two domain walls.  Furthermore,
since many unit cells are copied in this system, there are as 
many duplicates of the bands in the
enlarged system under consideration.  We study the bandgap 
precisely at the Dirac point as a function
of the fractional distance between the two domain 
walls on the torus $x$ (Fig. \ref{torus}) for a composite
unit cell consisting of $N=30$ unit cells copied along the $\bm{R}_{\bot}$ direction.  
When $x=0$ or
$x=1$, the two domain walls are at the same point, and this corresponds to a single domain with a
single Faraday coupling $\Lambda$.  For all other values of $x$, the ``unit cell'' comprises a 
two domain system
with non-equivalent lengths.  
In Fig. \ref{gapofx}, the gap between the two bands closest to the Dirac frequency
decays exponentially as a function of the distance between the two domain walls.  
We shall show that
the exponential decay in the gap corresponds to the localization of the edge modes along 
each domain wall.  
The small gap at intermediate values of $x$, when the two walls are far apart 
corresponds to the fact that 
each edge mode has a small amplitude, and therefore hardly 
mix with each other at those length scales.

\begin{figure}[htbp]
\begin{center}
\includegraphics[width=3.4in,keepaspectratio,clip]{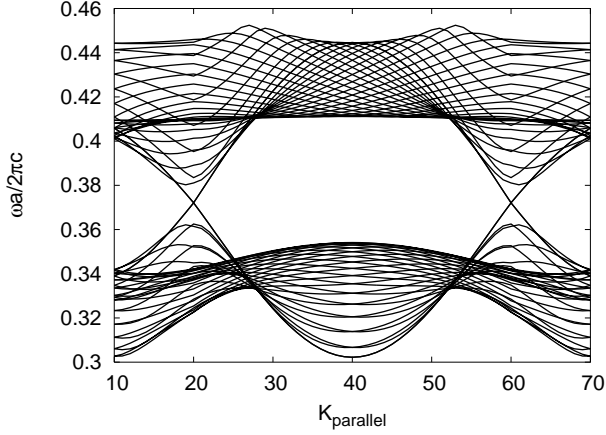}
\end{center}
\caption{The spectrum of the composite system consisting 30 copies of 
a single hexagonal unit cell duplicated along a direction $\bm{R}_{\bot}$.  
Both inversion and time-reversal symmetries are present, and the Dirac points are
clearly visible.  While the composite system has a spectrum containing many
bands, only two bands touch at the Dirac point.  The dispersion is computed in
$\bm{k}$ space along the direction \textit{parallel} to the wall.  }
\label{specA}
\end{figure}

\begin{figure}[htbp]
\begin{center}
\includegraphics[width=3.4in,keepaspectratio,clip]{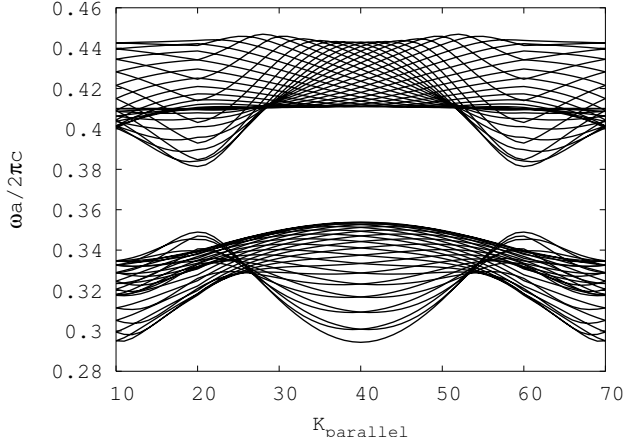}
\end{center}
\caption{The same system as above, but with broken time-reversal 
symmetry without a domain wall.  There is a single Faraday axis in 
the rods of the entire system.  }
\label{specB}
\end{figure}

\begin{figure}[htbp]
\begin{center}
\includegraphics[width=3.4in,keepaspectratio,clip]{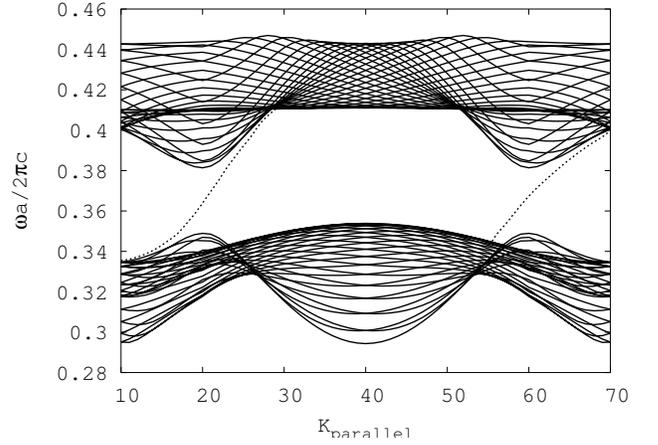}
\end{center}
\caption{Same system as above, but with a domain wall introduced corresponding
to maximum separation of the walls on the torus.  The two additional modes present
in the gap correspond to edge modes with a ``free photon'' linear dispersion along
the wall.  There are two modes, since across the domain wall, the Chern number of the band
just below the band gap changes by 2.}
\label{specC}
\end{figure}

When the domain wall is introduced, translational symmetry is still preserved along the direction
parallel to the wall, and the states of 
the composite system of 30 unit cells can be labeled by $\bm{k}_{\|}$, 
 Bloch vectors in the direction
parallel to the wall.
Figures \ref{specA}, \ref{specB},and \ref{specC} consist of a spectral series 
of a system without any broken time-reversal symmetry (Fig. \ref{specA}), with 
uniformly broken time-reversal 
symmetry (Fig \ref{specB}), and a domain wall configuration (Fig. \ref{specC}) 
for the 30 unit cell composite system.  
The bands 
are plotted along a trajectory in $\bm{k}$-space in the $\bm{k}_{\|}$ direction 
which contains the two distinct Brillouin zone corners.  It is clear that in the Domain wall, 
there are two additional modes formed in the bandgap that arose from the Faraday coupling.  
Since the domain walls are duplicated on the torus, the spectrum of edge modes will also be doubled; 
in Fig. \ref{specC}, only the two non-equivalent modes are shown.  
Each mode in the band gap has a free photon linear dispersion along the direction of the wall; 
moreover, both have positive group velocities, and therefore propagate \textit{unidirectionally}.  

To be certain, however, that these ``chiral'' modes are indeed localized near the interface, we
have numerically computed $\langle  u(\bm r)| \bm{B}^{-1}| u(\bm r) \rangle $, the 
 electromagnetic energy density (the $\bm{B}$ matrix, defined in section II, is not to be 
confused with the magnetic flux density), the photon probability density in real space.  We have
computed this quantity along with all the spectra of the composite system using the real space 
bandstructure algorithms described in Appendix \ref{app2}.  
As shown in Fig. \ref{gaussed}, the energy density is a gaussian function, peaked
at the position of the domain wall, decaying exponentially away from the wall.  
From this calculation,
we extract a localization also approximately 5 unit translations in the direction perpendicular
to the interface.  

We have therefore shown here using explicit numerical examples that photonic analogs of the
``chiral'' edge states of the integer quantum Hall effect can exist along domain walls of Hexagonal 
photonic systems with broken time-reversal symmetry.  We have studied the unphysical case in 
which such domain walls are abrupt changes in the axis of the Faraday coupling.  However, 
due to the topological nature of these modes, a smoother domain wall in which the Faraday 
axis slowly reverses over a length scale much larger than a unit cell dimension would also 
produce such modes.  The most important requirement for the existence of these modes, is that at 
some spatial location, the Faraday coupling is tuned across its critical value.  How this particular 
tuning is effected is irrelevant.  In the following section, after deriving the effective 
Hamiltonians for these modes, we shall consider a smoothly varying Faraday coupling, which 
corresponds to an exactly soluble system, and shall show the evolution of these modes as the 
smoothness of the Faraday coupling is varied towards the step function limit considered here.

\begin{figure}[htbp]
\begin{center}
\includegraphics[width=3.4in,keepaspectratio,clip]{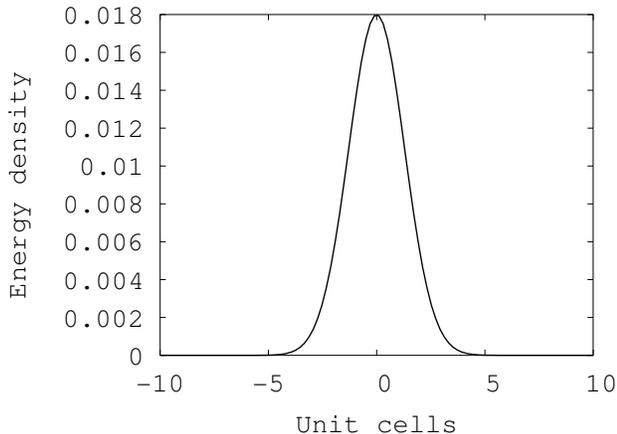}
\end{center}
\caption{The real-space electromagnetic energy density profile associated 
with the edge modes in Fig. \ref{specC} plotted as a function of the 
direction \emph{perpendicular} to the domain wall (and ``integrated'' over the direction parallel 
to the interface) and fit to a gaussian profile.  The integrated energy density depicted 
here plays the role of the photon probability density which confirms that light is confined 
to the interface. }
\label{gaussed}
\end{figure}

\section{Model Hamiltonian Approach}

The crucial feature exploited in the previous sections was the possibility of
tuning bandgaps at Dirac points by adding time-reversal breaking 
perturbations.  
Before adding these perturbations, the linear conical spectrum at these Dirac 
points are governed by two dimensional massless Dirac Hamiltonians, and 
time-reversal or inversion symmetry breaking perturbations contribute 
\textit{mass terms} to the Hamiltonian.  In this section, we shall construct 
these Dirac Hamiltonians starting from the Maxwell equations for 
two dimensional photonic systems with hexagonal symmetry.

To motivate Dirac Hamiltonians in photonic systems, we begin this section 
by considering  a ``nearly-free photon'' approach in which a
two dimensional ``free photon'' spectrum consisting of plane waves is 
perturbed 
by a weak periodic and hexagonal modulation of $\epsilon(\bm{r})$.  
Due to the underlying symmetry of the perturbation, 
the plane waves mix in a manner to generate 
Dirac points in the zone corners of this system.  
We then consider the effect adding time-reversal 
and inversion symmetry breaking perturbations in 
this system and derive an expression for the Dirac mass.  
Having motivated the Dirac points, we revert to our 
photon band problem and derive expressions for the Dirac 
mass in these systems.  

In analogy with the ``nearly-free electron'' approximation, we consider 
the photon propagation problem in the weak-coupling regime, in which the 
dielectric properties of the medium act as a weak perturbation.  We 
solve the Maxwell 
normal mode problem for Bloch state solutions, 
and work out corrections to the 
free photon dispersion relations in the Brillouin zone 
boundaries.  We shall assume 
continuous translational invariance in the z-direction, 
and study the propagation of electromagnetic waves in the x-y plane.

The free photon constitutive relations are 
isotropic and uniform in the plane: 

\begin{equation}
B_0 = \left( \begin{array}{cc}
\epsilon_0 & 0 \\
0 & \mu_0 \end{array} \right) .
\end{equation}
  
\begin{figure}[htbp]
\begin{center}
\includegraphics[width=2.0in,keepaspectratio,clip]{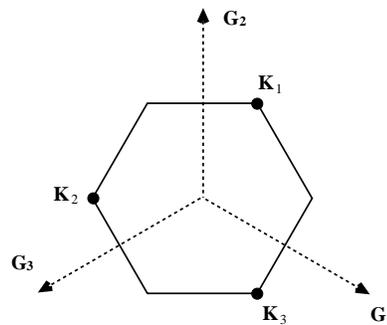}
\end{center}
\caption{In the weak coupling approach, the free photon TE mode plane waves
are perturbed by a periodic modulation in the permittivity.  The
plane wave frequency at the three equivalent zone corners ($\bm{K}_i$, 
i=1,2,3) is lifted by the permittivity in ``$\bm{k} \cdot \bm{p}$'' 
perturbation theory into a non-degenerate singlet and a degenerate 
doublet.}
\label{bz}
\end{figure}

We consider the ``free photon bands''  in the first
hexagonal Brillouin zone depicted in Fig. \ref{bz}.  
Let $\bm{G}_i$, $i=1,2,3$ be the three equal-length  
reciprocal lattice vectors each rotated $120^0$ with 
respect to one another.  The hexagonal zone corners 
correspond to the points $ \pm \bm{K}_i$, 
where $\bm{K}_1 = \left(\bm{G}_2 - \bm{G}_3 \right)/3$, 
etc., and $|\bm{K}| = |G|/\sqrt{3}$.  At each of the zone corners, 
the free-photon spectrum is six-fold degenerate with $\omega_0 = c_0 K$.  
In two dimensions, the modes decouple into TE $(E_x,E_y,H_z)$, 
and TM $(H_x,H_y,E_z)$ sets, and we shall focus only on the TE modes 
and consider the 3-fold TE mode symmetry at the zone corners (the TE and TM 
modes do not mix in 2 dimensions).  The eigenvalue equation for the free 
photon plane wave modes at the zone corners is 
$\bm A |\bm u_{0} \rangle = \omega_0 \bm B^{-1}_0 |\bm u_{0} \rangle $, 
or equivalently, 
$\bm B_0^{1/2}\bm A \bm B_0^{1/2} |\bm z_{0} \rangle = \omega_0 |\bm z_{0} \rangle$ and 
the states $|\bm z_{0} \rangle = \bm B_0^{-1/2} |\bm u_{0} \rangle $ satisfy $ 
\langle \bm z_0^{(\lambda)}|\bm z_0^{(\lambda ')} \rangle = \delta_{\lambda \lambda '}$.

Next, keeping the uniform isotropic permeability fixed, we add a weak
periodic perturbation to the permittivity  of the 
form 

\begin{equation}
\label{modulation}
\lambda B^{-1}_1 = \left( \begin{array}{cc}
\epsilon_0 \lambda V_G (\bm{r}) & 0 \\
0 & 0 \end{array} \right) ,
\end{equation}
with 

\begin{equation}
V_G(\bm{r}) = 2 \sum_{n=1}^3 \cos \left(\bm{G}_n \cdot \bm{r} \right) .
\end{equation}

After this perturbation is added, the TE and TM
modes no longer remain degenerate; while the TM modes remain 3-fold degenerate
at the zone corners at the frequency $\omega = c_0 |\bm{K}|$, the TE modes 
split apart into a singlet and a degenerate doublet.  We now determine the 
splitting to leading order in $\lambda$ with
within a weak-coupling ``nearly-free photon'' approach.  

With the periodic perturbation, the eigenvalue problem is

\begin{equation}
\bm A|\bm u \rangle = \omega \left(\bm B^{-1}_0 + \lambda \bm B^{-1}_1 \right) |\bm u 
\rangle ,
\end{equation}
which is equivalent to 
\begin{equation}
\bm B_0^{1/2} \left(\bm A - \lambda \omega \bm B^{-1}_1  \right) \bm B_0^{1/2} |\bm z 
\rangle = \left( \omega_0 + \delta  \omega \right) |\bm z \rangle .
\end{equation}

The energy splittings are worked out in degenerate perturbation theory (see 
Appendix \ref{app3}) as 

\begin{eqnarray*}
\frac{\delta \omega_n}{\omega_0} &=& -\lambda \langle \tilde{\bm z}_n | 
\bm B_0^{1/2} \bm B^{-1}_1 \bm B_0^{1/2} | \tilde{\bm z}_n \rangle \\
& =& - \lambda \langle \tilde{\bm u}_n |\bm B^{-1}_1| \tilde{\bm u}_n \rangle ,
\end{eqnarray*}
where $| \tilde{\bm z}_n \rangle $ are appropriate 
combinations of the three free photon plane-plane waves that diagonalizes the 
periodic potential.  These states are obtained by requiring them to be 
invariant under 3-fold rotations in the plane.  Instead of writing the 
fields in the coordinate basis, It is convenient to 
use a redundant basis of the three vectors 
$(e^{i\bm{K}_1\cdot \bm{r}},e^{i \bm{K}_2 \cdot \bm{r} },e^{ i \bm{K}_3 \cdot \bm{r}})$, 
with $\sum_n \bm{K}_n = 0$, 
and $\bm{K}_i \cdot \bm{K}_j = -K^2/2, i \neq j$.  
In this basis, the magnetic field of the TE modes is written as 
($\eta = e^{2 \pi i /3}$):

\begin{equation}
H^z_1 = \left(1,1,1 \right) ,
\end{equation}

\begin{equation}
H^z_2 = \left(1,\eta^*,\eta \right) ,
\end{equation}
and
\begin{equation}
H^z_3 = \left(1, \eta, \eta^* \right) . 
\end{equation}

The corresponding electric flux densities are easily obtained:

\begin{equation}
D^{\parallel}_1 = \frac{1}{\omega}\left(\hat{z} \times \bm{K}_1, \hat{z} 
\times \bm{K}_2, \hat{z} \times \bm{K}_3 \right) ,
\end{equation}

\begin{equation}
D^{\parallel}_2 = \frac{1}{\omega} \left(\hat{z} \times \bm{K}_1, 
\eta^* \hat{z} \times \bm{K}_2, \eta \hat{z} \times \bm{K}_3 \right) ,
\end{equation}

\begin{equation}
D^{\parallel}_3 = \frac{1}{\omega}\left(\hat{z} \times \bm{K}_1, 
\eta \hat{z} \times \bm{K}_2,\eta^* \hat{z} \times \bm{K}_3 \right) ,
\end{equation}
and
\begin{equation}
| \tilde{\bm z}_i \rangle = \left( \begin{array}{c}
E^{\parallel}_i \\
H^z_i \end{array}\right) .
\end{equation}

Clearly, these are the plane wave solutions that satisfy Maxwell 
equations and transform appropriately under 3-fold rotations in the plane.  
We are therefore led  
to the simple result that the splitting at the zone corners due 
to the mixing of the three plane waves is related 
to the integral over the unit-cell of the electric 
fields and the periodic potential, which is a traceless, 
real-symmetric $3 \times 3$ problem.  
It is easy to see that the problem is traceless 
because diagonal terms of the form $\langle u_i | B_1 | u_i \rangle$ vanish 
identically since $u_i$ are plane waves.  

To leading order in $\lambda$, the three photon bands split 
to form a singlet band at frequency 
$ \omega_0 = c_0|K| \left( 1 + \lambda/2  + O(\lambda^2) \right)$ 
and a degenerate doublet at frequency 

\begin{equation}
\omega_D = c_0|K| \left(1-\lambda/4 + O(\lambda^2) \right) .
\end{equation}

Exactly at the zone corners, the singlet and doublet states above 
diagonalize the perturbation in Eq.(\ref{modulation}).  
To leading order 
in $ \lambda $ and $\delta \bm{k} \equiv \bm{k} - 
\bm{K}_i$, the deviation in the Bloch vector from the zone corners, 
the states $ |\tilde{\bm z}_2 (\delta \bm{k})\rangle$ 
and $| \tilde{\bm z}_3 (\delta \bm{k})\rangle $,  
(where $ |\tilde{\bm z}_i(\delta \bm{k}) \rangle 
= \exp(i\delta \bm{k} \cdot \bm{r}) | \tilde{\bm z}_i \rangle)$, 
 which are degenerate at $\delta \bm{k}=0$  mix and split apart 
linearly as a function of $|\delta \bm{k}|$, forming a ``Dirac point''.  
To leading order, the Dirac point doublet does not mix with the singlet state 
$|\tilde{\bm z}_1 (\delta \bm{k} ) \rangle$.  
The effective 
Hamiltonian governing the spectrum of the doublet, 
to leading order in $\delta \bm{k}$
is a 2D massless Dirac equation:

\begin{equation}
\label{diracm0}
\omega_{\pm}(\delta \bm{k}) = \omega_D \pm v_D \left(\delta k_x \sigma^x + 
\delta k_y \sigma^y \right) ,
\end{equation}
where $v_D = c_0/2 + O(\lambda)$, and $ \sigma^i $ are the 
Pauli matrices written in the subspace of the doublet states.  The 
linear dispersion of the doublet in the neighborhood of the 
zone corners is immediately obtained by solving Eq.(\ref{diracm0}):

\begin{equation}
\omega = \omega_D \pm v_D |\delta \bm{k}| .
\end{equation}

The singlet band's frequency remains 
unchanged to leading order in $\delta \bm{k}$:  $\omega_0(\delta \bm{k}) 
= \omega_0 + O(|\delta \bm{k}|^2)$.  Thus, we have shown 
that the periodic modulation of the permittivity having 3-fold 
rotational symmetry gives rise to a quadratically dispersing singlet band 
and a ``Dirac point'' with linear dispersion.  

\begin{figure}[htbp]
\begin{center}
\includegraphics[width=3.0in,keepaspectratio]{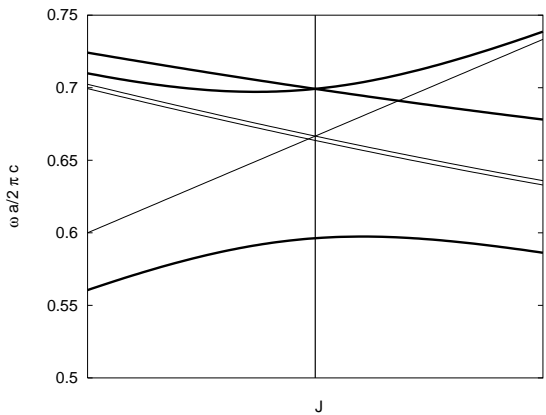}
\end{center}
\caption{Spectrum of photon dispersion in the vicinity of the 
zone corners.  We have arbitrarily set $\lambda < 0$ so that 
the singlet band has a lower frequency than the Doublet.  Free 
photon spectra are given by dashed lines.  Away from the zone corners, 
the free spectrum is not affected to leading order in $\lambda$.}
\end{figure}

Next, we add a Faraday term, with an axis normal to the $xy$-plane,
 to the permittivity tensor 
$ \epsilon^{xy} = - \epsilon^{yx} = i \epsilon_0 \eta (\bm{r}, \omega)$, 
where

\begin{equation}
\eta(\bm{r},\omega)= \eta_0 (\omega) + \eta_1 (\omega) V_G(\bm{r}) .
\end{equation}

Both $\eta_0 (\omega)$ and $\eta_1 (\omega)$ are odd functions of $\omega$.  
In the limit that the Faraday coupling is much weaker in strength than 
the periodic modulation, $|\eta_0|$,  $| \eta_1| \ll |\lambda| \ll 1 $, the 
mixing between the non-degenerate singlet state and the doublet remains 
negligible, and the energy of the singlet state is unaffected by the Faraday 
perturbation.  However, the doublet states split apart at the Dirac point.  
Using the expression for the Dirac point splitting, derived in the 
Appendix \ref{app3}, we find that the splitting of the doublet at the zone corner is 
given by 
\begin{equation}
\omega_{\pm} - \omega_D = \pm v_D \kappa , \quad \kappa = |K| \left(\frac{3}{2} 
\eta_1 (\omega_D) - 3 \lambda \eta_0 (\omega_D) \right) .
\end{equation}
Away from the Dirac point (but still close enough to 
the zone corners so that the ``nearly-free photon'' 
approximation for the three plane wave states remains valid), 
the doublet bands acquire a dispersion 
\begin{equation}
\label{splitting} 
\omega = \omega_D \pm v_D \left(|\delta \bm k|^2 + \kappa^2 \right)^{1/2} ,
\end{equation}
which is the spectrum of a 2D massive Dirac Hamiltonian:
\begin{equation}
\omega_{\pm}(\delta \bm k) = \omega_D \pm v_D \left( \delta k_x \sigma^x 
+ \delta k_y \sigma^y + \kappa \sigma^z \right) .
\end{equation}

The Dirac points that occur in the ``nearly-free photon'' approximation 
are not isolated points of degeneracy, since, away from the zone 
corners, the two bands which formed the Dirac point merge together to 
resume their original free-photon form.  Consequently, the type of modes 
studied in the previous section cannot be reproduced using this type of 
weak-coupling expansion.  

However, we can gain understanding by suppose that we have the exact 
solutions of the electromagnetic 
Bloch states and eigenfrequencies of a system containing isolated Dirac 
points, such as the one studied numerically in section IV.  We can use 
precisely the same  weak Faraday coupling approximation to 
work out the splitting of the Dirac point 
with a Faraday term.  Assuming we are given example photonic bandstructures 
of long hexagonal systems with $k_z = 0$, which contain only 
isolated Dirac points, a weak Faraday coupling would split apart the 
bands that formed the Dirac point, and the splitting is identical 
to that in (\ref{splitting}).  Suppose that the two bands having a 
Dirac point, otherwise form a PBG with a gap $\Delta \gg v_D \kappa$ (as in 
the case of the numerical example given in the previous section.   
In this case, since the Faraday term removes 
all points of degeneracy, the now non-degenerate bands have a well-defined 
Berry curvature field
\begin{equation}
\mathcal{F}_{\pm}(\delta \bm k) = \pm \frac{1}{2} \kappa 
\left( |\delta \bm k|^2 + \kappa^2 \right)^{-3/2},
\end{equation}
which decays rapidly away from the Dirac point, and contributes a total 
integrated Berry curvature of $\pm \pi$.   Since there are two 
non-equivalent Dirac points in the hexagonal geometry under 
consideration, the net Berry curvature of the system is the sum of the 
contributions from each Dirac point.  If, as in the case under consideration, 
spatial inversion symmetry is preserved, but time-reversal symmetry is broken,
the Berry curvature fields at each Dirac point of a given band add, giving 
total Chern numbers $\pm 1$ for each of the split bands.  However, 
if time-reversal symmetry were preserved, and inversion symmetry breaking 
caused the gap to open, the Berry curvature field of each Dirac point 
for a given band are equal in magnitude but opposite in sign, and the 
Chern number would vanish.  

As before, to get unidirectional edge modes of light in this system, the 
Faraday coupling must be tuned across its critical value $\eta (\bm r, \omega) = 0$.  
To do this, we consider a Faraday coupling that varies slowly 
and adiabatically in space, we shall assume negligible frequency dependence 
of the Faraday coupling, and we shall parameterize the local value of the 
Faraday coupling by a smoothly varying function $\kappa (\bm r)$, which 
is positive in some regions 
and negative in other regions of the 2D plane perpendicular to the 
cylindrical axis of the hexagonal array of rods.  Due to the adiabatic 
variation of $\kappa( \bm r)$, each point in space is characterized by 
a local bandstructure problem, and the splitting at the Dirac point is 
given again by the expression in (\ref{splitting}), but with the local 
value of $\kappa$.  In this limit, the smooth variation of $\kappa(\bm r)$ 
leads to a 2D Dirac Hamiltonian with a adiabatically spatially varying 
mass gap.  At all points where $\kappa (\bm r) = 0$, the local bandstructure 
in the vicinity of the Dirac point is the massless 2D Dirac Hamiltonian; 
provided that $|\kappa ( \bm r)| \ll \Delta$, the PBG,  
the spectrum far away from the Dirac points is unaffected by 
$\kappa(\bm r)$.  In what 
follows, we assume that when $\kappa = 0$, our bandstructure 
contains Dirac points 
which are formed by two isolated bands 
in a PBG region having no other points of degeneracy.  

We neglect the mixing between modes at different Dirac points, and consider 
the situation in which $\kappa (\bm r)$ vanishes along a single line ( $ x = 0$
 for instance), and we assume translational invariance along the 
direction parallel to the interface ($y-$ direction).  As before, we 
consider the degenerate perturbation problem of the normal modes close 
to the Dirac point.  Now, however, the coefficients of the degenerate 
solutions of the Maxwell equations are spatially varying quantities.  Let 
$| u_{\sigma} (\pm \bm k _D) \rangle$, $\sigma = \pm$, be the degenerate 
solutions (i.e. the periodic parts of the photon Bloch state wave 
functions) at a pair of Dirac points when $\kappa = 0$.  With the local 
variation, we take spatially varying linear combination of these Bloch states 
\begin{equation}
u(\bm k _D, \bm r) = \sum_{\sigma, \pm} \psi_{\sigma}(\bm r) 
\exp\left(\pm i \bm k _D \cdot \bm r \right) u_{\sigma}(\pm \bm k _D, \bm r) ,
\end{equation}
and arrive at the fact that the local value of the splitting of the two 
bands at $\bm k_D$ is 
\begin{equation}
\omega_+(\bm k_D) - \omega_- (\bm k_D) = 2\kappa (\bm r)  .
\end{equation}

In the neighborhood of the Dirac point, the degenerate perturbation 
problem gives us a 2D massive Dirac Hamiltonian, with $\delta k_x$ 
replaced by the operator $-i \nabla_x$ in the position representation, 
since translation symmetry in the x-direction 
is broken by $\kappa (x)$.  We thus obtain an expression of the form 
$v_D \hat{K} | \psi \rangle = \delta \omega | \psi \rangle$, and 
\begin{equation}
\hat{K} = -i \bm \sigma^x \nabla_x + \delta k _{\parallel} \bm \sigma^y 
+ \kappa(x) \bm \sigma^z .
\end{equation}
The Bloch vector in the y-direction, which remains conserved due to the 
preservation of translation invariance along this direction, is $k_{D y} + 
\delta k_{\parallel}$.  

For the particular choice of $\kappa(x)= \kappa^{\infty}\tanh(x/\xi)$, 
$\xi > 0$, (where $\kappa^{\infty}$ is the asymptotic value of the Dirac point 
splitting at distances $\gg \xi$ from the interface), 
the problem is exactly solvable, since the Dirac Hamiltonian $\hat{K}$,  
when squared, becomes a 1D Schr\"odinger Hamiltonian $\hat{K}^2$ corresponding 
to the integrable Poschl-Teller Hamiltonian \cite{landauIII}.  

To see how this comes about, we explicitly work out the operator $\hat{K}^2$,
making use of the anti-commuting property of the 
Pauli matrices $\lbrace \sigma^a, \sigma^b \rbrace = 2 \delta^{ab}$ :  
\begin{equation}
\hat{K}^2 - \delta k_{\parallel}^2 = - \nabla_x^2 + \kappa(x)^2 - \bm \sigma^y \kappa' .
\end{equation}
The spatially varying Dirac mass term that changed sign across the interface 
becomes a ``potential well'' with bound states given by\cite{landauIII}

\begin{eqnarray}
\omega_0 (\delta k_{\parallel}) &=&\omega_D + s_{\kappa} v_D 
\delta k_{\parallel}, \quad s_{\kappa} = sgn(\kappa^{\infty})\\
\omega_{n \pm} &=& \omega_D \pm v_D \left(\delta k_{\parallel}^2 + 
\kappa_n^2 \right)^{1/2}, \quad n>0 .
\end{eqnarray}
where $|\kappa_n = 2n|\kappa^{\infty}|/ \xi$, $n < |\kappa^{\infty} \xi /2$.  
In the $n=0$ mode, light propagates unidirectionally, with velocity $v_D$, 
in the direction parallel to the wall.  All other bound modes are 
bidirectional modes.  The numerical example of a Dirac 
mass studied in the previous section that changed sign abruptly, as 
a step function, has the 1D Schrodinger problem in an attractive 
delta function potential as its square.  Consequently, as we have seen, 
the model permitted for only a single bound state, corresponding to the 
unidirectional mode.

\begin{figure}[htbp]
\begin{center}
\includegraphics[width=3.0in,keepaspectratio]{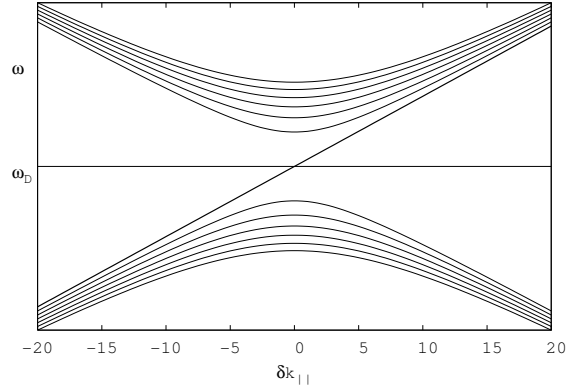}
\end{center}
\caption{Spectrum of the integrable Poschl-Teller model.  With the 
exception of the zero mode, all bound states correspond to 
bi-directionally propagating modes localized at the interface where 
the function $\kappa(\bm r) = 0$.  The zero mode, on the other hand, 
is unbalanced, and furthermore, it corresponds to unidirectional 
propagation.  }
\label{poschl-teller}
\end{figure}

For the generic case, the second order differential equation for the 
$n > 0$ bound states can not be solved analytically.  However, a formal 
solution for the zero mode eigenfrequency can be obtained, as it is 
obtained by solving a \emph{first order} equation, as we now discuss. 
Starting from the Dirac equation for the more general case
\begin{equation}
v_D \left( -i \bm \sigma^x \nabla_x + -i \bm \sigma^y \nabla^y + \kappa(x) 
\bm \sigma^z \right) | \psi_{\pm} \rangle = \delta \omega | \psi_{\pm} \rangle ,
\end{equation}
by definition, the ``zero mode'' has the free photon dispersion along 
the direction parallel to the wall, which implies that the function 
$|\psi \rangle \propto \exp(i \delta k_{\parallel} y)$.  We are thus 
left with the equation 
\begin{equation}
\left( -i \bm \sigma^x \nabla_x + 
\kappa(x) \bm \sigma^z \right)|\psi_{\pm} \rangle  = 0 .
\end{equation}
Multiplying both sides with $\bm \sigma^x$, we arrive at the following 
first order differential equation :
\begin{equation}
\left(\nabla_x + \kappa(x) \bm \sigma^y \right) = 0 ,
\end{equation}
which has as it's formal solution
\begin{equation}
|\psi_{\pm} \rangle = \exp\left( i \delta k_{\parallel} y + \alpha \int^x dx'
\kappa(x') \right) | \phi_{\pm} (\alpha) \rangle ,
\end{equation}
where $\bm \sigma^y | \phi_{\pm}(\alpha) \rangle = \alpha | \phi_{\pm} 
(\alpha) \rangle$.  
Although there are formally two solutions for the zero mode, corresponding to 
$\alpha = \pm 1$, only one can occur; the other is not normalizable and thus 
cannot represent a physically observable state.

\section{Semiclassical Analysis}
Now let the ``Dirac mass'' term that opens the photonic band gap
be a slowly varying function $\kappa(x)$ that changes monotonically
(and analytically) from $-k_0$ at $x$ = $-\infty$ to $k_0$ at $x$ = $+\infty$.
The photonic spectrum of 
modes with
wavenumbers $\bm k$ = $\bm k_D+\delta\bm k$
near the ``Dirac point'' $\bm k_D$, and which become
doubly-degenerate at $\bm k_D$, is an adiabatic function
of $x$:
\begin{eqnarray}
\omega(x, \delta k_x,\delta k_y) &=& \omega_D
\pm v_D\left ( \delta k_y^2 + k(x,\delta k_x)^2 \right )^{1/2},
\nonumber \\
k(x,\delta k_x)^2 &=& \delta k_x^2 + \kappa(x)^2 ,
\end{eqnarray}
where $v_D > 0$ is the ``Dirac speed''. 
For $k(x,\delta k_x)^2 < k_0^2$, the modes are evanescent as
$x \rightarrow \pm \infty$, so are localized on the wall.
In the $x-\delta k_x$ plane, the contours of constant
$k(x,\delta k_x)^2   < k_0^2$ are simple closed curves, enclosing a finite
dimensionless area $\phi(k^2)$, given by
\begin{eqnarray}
\phi(k^2) &=& 2\int_{x_-}^{x_+}dx \left ( k^2 - \kappa(x)^2\right )^{1/2} , 
\end{eqnarray} 
where $x_-(k^2)$  $ <$  $x_+(k^2)$ are the two ``turning point''
solutions of $\kappa(x_{\pm})^2$ = $k^2$ . 
Since $\kappa(x)$ is assumed to be monotonic, this can be written as
\begin{eqnarray}
\phi(k^2) &=& 2\int_{0}^{|k|} dy 
\left ( k^2 - y^2\right )^{1/2}
\left ( \frac{1}{\kappa'_+(y^2)} + \frac{1}{\kappa'_-(y^2)} \right ),
\nonumber \\
\kappa'_{\pm}(k^2) &\equiv& \left .\frac{d\kappa}{dx} \right |_{x_{\pm}(k^2)},
\end{eqnarray}
Note that this transformation has turned $\phi(k^2)$ into
a \textit{signed} area, where ${\rm sgn}(\phi)$ =
${\rm sgn}(k_0)$, which is indeed the correct form 
(the function $\phi(k^2)$ vanishes
as $k_0 \rightarrow 0$, when its domain $k^2 \le k_0^2$ shrinks to zero).
In the limit $k^2 \rightarrow 0$, $x_{\pm}(k^2) \rightarrow x_0$, the
formal location of the interface.  Then
$\kappa'_{\pm}(k^2)$ $\rightarrow$ $\kappa'(x_0)$, and
 $\phi(k^2)$ vanishes as  
\begin{equation}
\phi(k^2) \rightarrow \frac{\pi k^2}{\kappa'(x_0)},
\quad (k/k_0)^2 \rightarrow 0.
\label{limiting}
\end{equation}

It is very instructive to examine the special case
\begin{equation}
\kappa(x) = k_0\tanh (\alpha (x-x_0) ),
\label{tanh}
\end{equation}
which is integrable.
In this case,
\begin{eqnarray}
&&\kappa'(x) = \alpha k_0 {\rm sech}^2(\alpha(x-x_0)), \\
&&k_0{\rm sech}^2(\alpha (x_{\pm}(k^2)-x_0))
= \frac{k_0^2 - k^2}{k_0}.
\end{eqnarray}
Thus the explicit dependence on $x_{\pm}(k^2)$ can be
eliminated, and
\begin{equation}
\kappa'_{\pm}(k^2) = 
\alpha  \left ( \frac{k_0^2 - k^2}{k_0}
\right ).
\end{equation}
This make the integral for $\phi(k^2)$ trivial (it becomes
expressible in terms of a simple Hilbert transform), and the
asymptotic small-$k^2$ form (\ref{limiting}) remains
valid for \textit{all} values of $k^2$ in the domain
of the function:
\begin{equation}
\phi (k^2) = \frac{\pi k^2}
{\alpha k_0},
\quad k^2 \le k_0^2.
\end{equation}
Then the frequency of the interface mode
with wavenumber $\delta k_y$ = $\delta k_{\parallel}$ along the
interface can be expressed as
\begin{eqnarray}
\omega(\delta k_{\parallel},\phi) &=&
\omega_D \pm v_D
\left (\delta k_{\parallel}^2 + \kappa_{\perp}^2(\phi)\right )^{1/2},
\nonumber \\
\kappa^2_{\perp}(\phi) &\equiv& |\alpha k_0 \phi| /\pi.
\end{eqnarray}
A standard ``semiclassical'' analysis of
interference effects on a light ray
trapped in a ``waveguide'' 
at an interface would conclude that the ``quantized''
values of $\phi$ corresponding to interface modes 
were
\begin{equation}
\phi_n = 2\pi n + \gamma,
\end{equation}
where $\gamma$ is a ``Maslov phase'', usually $\pi$.
In this case, comparison with the exact solution
of the integrable problem 
confirms that this problem instead has a vanishing
``Maslov phase'' $\gamma$ =  0.   This can be
attributed to  an underlying 
``$Z_2$'' Berry phase factor of $-1$ (Berry phase of $\pi$)
for orbiting around the degeneracy point
at $(x-x_0,k_x)$ =$(0,0)$.

We then conclude that the   interface
modes at a slowly-varying interface are in general
given (for small $\delta k_{\parallel}$) by
\begin{eqnarray}
\omega_0(\delta k_{\parallel}) &=& \omega_D
+  v_D{\rm sgn (k_0)}\delta k_{\parallel} ,\nonumber  \\ 
\omega_{n\pm} (\delta k_{\parallel}) 
&=& \omega_D \pm v_D\left (\delta k_{\parallel}^2 + k_n^2\right )^{1/2},
n \ge 1 ,\nonumber \\
\phi (k_n^2) &=& 2\pi n, \quad k_n^2 \le k_0^2.
\end{eqnarray}
The unidirectional ``zero mode'' persists however sharp the interface
is; the bidirectional
modes with $n\ge 1$ must obey $2\pi n < \phi(k_0^2)$, which
has fewer and fewer (and eventually no) solutions as
the width of the interface region shrinks.
In the special case of the integrable model (\ref{tanh}),
this spectrum is exact for small $\delta k_{\parallel}$
without any condition that the wall is slowly varying.

\section{Discussion}

The occurrence of zero energy modes in the 2D Dirac Hamiltonian is well known 
and represents the simplest example of a phenomenon known as the ``Chiral 
anomaly''.  The crucial feature, namely, the occurrence of interfaces 
where the Dirac mass gap changes sign, corresponds to tuning our photon 
band problem across a critical point using a Faraday effect.  

We have shown that analogs of quantum Hall effect edge modes can exist 
in photonic crystals whose band gaps can be tuned by a Faraday coupling.  
The crucial new feature we present here here is that photonic systems can 
have bands with non-trivial topological properties including non-zero 
Chern invariants.  These in turn can be varied in a controlled manner 
to yield unidirectional (``chiral'') edge modes.  The edge modes are robust 
against elastic back-scattering since they are states which are protected 
by the underlying 2D band structure topology.  However, they are not 
robust against photon number non-conserving processes, such as 
absorption and other non-linear effects.  We believe that this 
could be an entirely new direction in ``photonic band structure 
engineering'' due to the absence of scattering at bends and 
imperfections in the channel.  

\begin{figure}[htbp]
\begin{center}
\includegraphics[width=3.4in,keepaspectratio,clip]{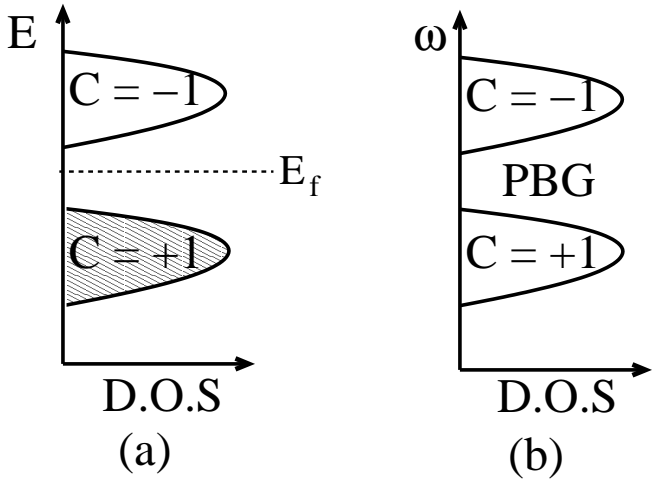}
\end{center}
\caption{We have shown in this paper that although the photon bands (b) 
cannot be filled as in the electronic case (a), they can have no 
analog of the bulk quantum Hall effect.  However, the Chern number is 
a topological invariant of Bloch states independent of the constituents.  
With the Faraday term, we are able to tune the system such that the total 
Chern number below a photonic band gap changes across in interface, which 
gives rise to unidirectionally propagating edge modes of photons localized 
at the interface.  These modes are direct analogues of the ``Chiral'' 
edge modes of electronic systems which occur at interfaces between two 
regions having different total Chern invariants below the Fermi level (i.e. 
with different Hall conductances).  }
\end{figure}

A practical realization of such one-way transmission channels 
in photonics will necessarily have to deal with the problem of 
finding a magneto-optic material with a strong enough Faraday 
effect to confine the light close to the interface.  Furthermore, 
in a practical design, the problem of TE/TM mode mixing when light 
is confined in the direction perpendicular to the 2d system will have 
to be addressed.  A practical design could, for instance, make use of PBG 
materials to confine light in the z-direction.  
Although there are many obstacles to the realization 
of such interesting effects in photonics, none of them are fundamental, 
and we believe that these 
unidirectional channels could have potentially useful technological 
applications which could in principle be realized someday through
``bandstructure engineering''.

\begin{acknowledgments}
This work was supported in part by the U. S. National Science Foundation 
(under MRSEC Grant No. DMR-0213706 at the Princeton Center for 
Complex Materials).  Part of this work was carried out at the Kavli Institute 
for Theoretical Physics, Santa Barbara, with support from KITP's NSF 
Grant No. PHY99-07949.

\end{acknowledgments}

\begin{appendix}

\section{\label{app1}Frequency dependence of the dielectric media}
In this section, we shall provide the details of the generalization 
of the normal mode problem to include the frequency-dependent response 
of the media outlined in section II.    
We shall couple 
the electromagnetic fields to harmonic oscillator degrees of freedom of the medium.  
Defining $\phi_{i \sigma}$ and $\pi_{i \sigma}$ ($i=1,\cdots N$, $\sigma = \epsilon,\mu $)
 to be a set of N independent canonically conjugate oscillator coordinates 
and momenta respectively, which represent internal polarization and magnetization modes, 
we consider the total Hamiltonian 

\begin{equation}
\label{total Hamiltonian}
H = H^{em} + \sum_{\sigma} H^{\sigma}  \ , (\sigma = \epsilon, \mu) ,
\end{equation}
where, for instance, 

\begin{eqnarray*}
H^{\epsilon} = \sum_i D^a \left( \alpha_{i \epsilon}(\bm{r})^a \pi_{i \epsilon}(\bm{r}) + 
\beta_{i \epsilon}^a(\bm{r}) \phi_{i \epsilon}(\bm{r}) \right) \\
+ \frac{1}{2}\sum_i \omega_{i \epsilon} \left( \pi_{i \epsilon}(\bm{r})^2 + 
\phi_{i \epsilon}(\bm{r})^2  \right) .
\end{eqnarray*} 

The first term above represents the local coupling between the electric fluxes and the polarization 
modes, whereas the second term represents the energy of the oscillators themselves.  A 
similar equation exists for the magnetization degrees of freedom coupled with the magnetic fluxes.  
The Hamiltonian, as stated in Eq.(\ref{total Hamiltonian}), is real-symmetric and positive-definite, 
and therefore, its eigenvalues are real.  
The electric and magnetic fields are obtained by varying the Hamiltonian with respect to the 
associated flux densities: $E_a(\bm{r})$ = 
$\delta \mathcal{H}/\delta D^a (\bm{r})$, 
$H_a(\bm{r})$ = $\delta \mathcal{H}/\delta B^a (\bm{r})$

\begin{equation}
\label{self-consistent1}
E_a( \bm{r}) = \epsilon^{-1}_{ab}(\bm{r})D^b(\bm{r}) + 
\sum_n \left(\alpha_{n \epsilon}^a (\bm{r})\phi_{n \epsilon}(\bm{r}) + 
\beta_{n \epsilon}^a(\bm{r}) \pi_{n \epsilon}(\bm{r}) \right) ,
\end{equation}
and similarly for the field $H_a$.  The time-evolution of the oscillator modes 
are obtained from the Hamilton equations of motion (letting 
$ \partial_t \phi_{n \sigma} = -i \omega \phi_{n \sigma}$, etc)

\begin{eqnarray}
-i \omega \phi_{n \epsilon}(\bm{r}) = \frac{\delta \mathcal{H}}{\delta \pi_{n \epsilon}(\bm{r})} =
 \omega_{n \epsilon} \pi_n(\bm{r}) + \beta^a_n(\bm{r}) D^a(\bm{r}) \\
 i \omega \pi_{n \epsilon}(\bm{r}) = \frac{\delta \mathcal{H}}{\delta \phi_{n \epsilon}(\bm{r})} = 
\omega_{n \epsilon} \phi_{n \epsilon}(\bm{r}) + \alpha_{n \epsilon}^a(\bm{r}) 
D^a(\bm{r}) .
\end{eqnarray}

We invert this equation to solve for the oscillator coordinates and momenta in terms of the fluxes:

\begin{equation}
\label{self-consistent2}
\left( \begin{array}{c}
\phi_{n \epsilon}(\bm{r}) \\
\pi_{n \epsilon}(\bm{r}) \end{array} \right) = \frac{1}{\omega^2 - \omega_n^2 } \left( \begin{array}{cc}
\omega_n  & i \omega \\
-i \omega & \omega_n \end{array} \right) \left( \begin{array}{c}
\alpha_{n \epsilon}^a(\bm{r}) \\
\beta_{n \epsilon}^a(\bm{r}) \end{array} \right) D^a(\bm{r}) .
\end{equation}

By substituting Eq.(\ref{self-consistent2}) into the 
expression for the electric field (\ref{self-consistent1}), we obtain 
a correction $\delta \epsilon^{-1}_{ab}(\bm{r} \ \omega)$ to 
the permittivity tensor  coming 
from the oscillator modes:

\begin{equation}
\delta \epsilon^{-1}_{ab}(\bm{r} \ \omega) = 
\sum_n \left(\frac{\Gamma^{\epsilon}_{ab}(\bm{r})(\omega+\omega_n)-
\Gamma^{*\epsilon}_{ab}(\bm{r})(\omega-\omega_n)}{\omega^2 - \omega_n^2}  \right) ,
\end{equation}
where 

\begin{equation}
\Gamma_{\epsilon}^{ab}(\bm{r}) = \left(\alpha_{n \epsilon}^a(\bm{r}) - i 
\beta_{n \epsilon}^a(\bm{r}) \right) \left( \alpha_{n \epsilon}^b(\bm{r}) + i \beta_{n 
\epsilon}^b(\bm{r}) \right) .
\end{equation}

Finally, the correction term above to the permittivity  
is expressed in Kramers-Kr\"onig form as 
\begin{equation}
\delta \epsilon^{-1}_{ab}(\bm{r}, \ \omega) = \sum_{n} \left( 
\frac{\Gamma^{\epsilon}_{ab}(\bm{r})}{\omega - \omega_n} - \frac{\Gamma^{\epsilon 
*}_{ab}(\bm{r}) }{\omega+ \omega_n } \right) .
\end{equation}

The same formal manipulations occur in the frequency dependence of the 
magnetization modes; in the end, the constitutive relations are given by a 
tensor $B(\bm{r} , \ \omega)$ defined by
\begin{equation}
B(\bm{r}, \omega) = \left(\begin{array}{cc}
\epsilon^{-1}(\bm{r}, \ \omega) & 0 \\
0 & \mu^{-1}(\bm{r}, \ \omega) \end{array} \right), 
\end{equation}
which is written in Kramers-Kronig form as:

\begin{equation}
\label{binv}
B_{ab}(\bm{r}, \ \omega) = S_{ab}(\bm{r}) + \sum_{n} \left( \frac{\Gamma_{ab}(\bm{r})}{\omega - 
\omega_n} - \frac{\Gamma_{ab}^*(\bm{r})}{\omega + \omega_n} \right) .
\end{equation}

The first term, $S_{ab}(\bm{r}) = \lim_{\omega \rightarrow \infty} B(\bm{r}, \ \omega)$ is the 
same tensor defining the Hamiltonian in Eq.(\ref{ham0}).
In the zero frequency limit, 

\begin{equation}
\label{binv0}
B_{ab}(\bm{r}, \ 0) = S_{ab}(\bm{r}) - \sum_n\left(\frac{\Gamma_{ab}(\bm{r})+
\Gamma_{ab}^*(\bm{r})}{\omega_n} \right).
\end{equation}

Stability of the medium imposes the following important constraint:
\begin{equation}
B(\bm{r}, \ 0) > 0 .
\end{equation}

Eliminating $S_{ab}$ in Eq.(\ref{binv}) using Eq.(\ref{binv0}), we get
$$
\delta B(\omega) = \sum_n \left[ \Gamma \left(\frac{\omega}{\omega_n(\omega-\omega_n)} \right) 
+ \Gamma^* \left( \frac{\omega}{\omega_n(\omega+\omega_n)} \right) \right] $$
where $\delta B(\omega) = B(\omega) - B(0)$.  
Whereas $B(\omega)$ is not a positive-definite matrix, the quantity which 
is guaranteed to be positive-definite in lossless frequency ranges is 
\begin{equation}
\label{binvtilde}
\tilde{B}(\omega) = B(\omega) - \omega \frac{\partial }{\partial \omega} B (\omega) > 0 
,
\end{equation}
because 

\begin{widetext}
\begin{equation}
\tilde{B} (\omega) = B (0) + \sum_n \frac{1}{\omega_n} 
\left[ \Gamma_n \left(\frac{\omega}{\omega-\omega_n} \right)^2 + 
\Gamma^*_n \left( \frac{\omega}{\omega+\omega_n} \right)^2 \right] ,
\end{equation}
\end{widetext}
and $\Gamma_n$, $\Gamma_n^*$, and $B(0)$ are all positive-definite 
tensors.  

Although $B(\omega)$ is not positive-definite, we will be interested 
in cases where 
\begin{equation}
\label{detbinv}
Det(B(\omega)) = 0 . 
\end{equation}

When this condition is satisfied and $B(\omega)$ has no zero modes 
corresponding to metallic conditions, there is a well defined inverse 
tensor $B^{-1}(\omega)$ 
\begin{equation}
B^{-1}(\bm{r}, \ \omega) = \left( \begin{array}{cc}
\epsilon(\bm{r}, \ \omega) & 0 \\
0 & \mu(\bm{r}, \ \omega) \end{array} \right) .
\end{equation} 
From the stability condition stated for $B(\omega)$, there 
exists a similar condition for $B^{-1}(\omega)$:
\[ B - \omega \frac{\partial}{\partial \omega} B \\
  = B \left(B^{-1} + \omega \frac{\partial}{\partial \omega} B^{-1} \right)B > 0, \]
where we have made use of $B^{-1}B= 1$ and 
$\partial / \partial \omega \left( B^{-1}B \right) = 0$.  
Supplementing the inequality above with the condition in Eq.(\ref{detbinv}), 
we obtain 
\begin{equation}
\label{stability}
\tilde{B} ^{-1}(\omega) \equiv \frac{\partial}{\partial \omega} \left( 
\omega B^{-1}(\omega) \right) > 0 .
\end{equation}

The eigenvalue problem is solved for each value 
of the bloch vector $\bm{k}$ in the first Brillouin zone, and 
The formal strategy for obtaining the energy eigenvalues
is to solve $\bm{A}|\bm{u}_n (\bm{k}) \rangle = 
\omega_n (\bm{k}) \bm{B}^{-1} (\omega(\bm{k})) |\bm{u}_n(\bm{k}) \rangle$, 
and then to \textit{vary} $\omega$ until it coincides with a frequency of an eigenmode.  
The stability condition (see Eq.(\ref{stability})) guarantees that such a prescription 
enables us to find the entire spectrum in a lossless range of real frequencies, where 
$\tilde{B}^{-1}$ is Hermitian.    

Indeed, if we consider for the moment the Hermitian problem 
\begin{equation}
\left( \bm{A} - \omega \bm{B}^{-1} (\omega) \right) 
| \bm{u}_n \rangle = \lambda_n (\omega) | \bm{u}_n \rangle ,
\end{equation}
and vary $\omega$ to find the \textit{zero modes}
\begin{equation}
\lambda_n (\omega) = 0 ,
\end{equation}
the stability of such a prescription is guaranteed only if 
\begin{equation}
\frac{\partial \lambda_n}{\partial \omega} < 0 ,
\end{equation}
so that the eigenvalues are monotonically decreasing functions of $\omega$.  
But from first order perturbation theory, we know that the requirement 
above is satisfied only if 

\begin{equation}
 \langle \bm{u}_n | \frac{d}{d \omega} \left( \omega B^{-1}(\omega)\right) 
| \bm{u}_n \rangle > 0 ,
\end{equation}
which is precisely equivalent to the condition in Eq.( \ref{stability}).  

When we eliminate the internal oscillator (polariton) modes and 
explicitly substitute
the expressions in Eq.(\ref{self-consistent2}) into the total Hamiltonian, Eq.
(\ref{total Hamiltonian}), we obtain the following quadratic form that involves 
only the electromagnetic flux densities:

\begin{equation}
\label{effective Hamiltonian}
H = \frac{1}{2}\sum_{ij}\tilde{B}^{-1}_{ij}(\omega)(u_i)^*u_j .
\end{equation}

Our result can be summarized as follows.  We begin with our total Hamiltonian,
 Eq.(\ref{total Hamiltonian}), which can be written as a positive-definite 
real-symmetric matrix whose states live in an enlarged Hilbert 
space containing electromagnetic flux densities and internal oscillator modes. 
When we ``integrate out'' the non-resonant internal oscillator modes 
of the media, we are left with a set of effective constitutive relations 
of the form 
\begin{equation}
v_{i \lambda} = \sum_j B_{ij}(\omega_{\lambda})u_{j \lambda}^+ 
e^{i\omega_{\lambda}t} + c.c. ,
\end{equation}
and an effective Hamiltonian (which represents the conserved 
time-averaged energy density of the electromagnetic fields as well 
as the oscillator modes) that involves a \textit{different} tensor 
$\tilde{B}_{ij}(\omega)$ given in \ref{effective Hamiltonian}.  
Using the relation in Eq.(\ref{binvtilde}), we can equivalently 
write the Hamiltonian as 
\begin{equation}
H = \frac{1}{2} \sum_{ij} \tilde{B}_{ij} (v_i)^* v_j .
\end{equation}

For the case of generalized frequency dependence considered here, the 
normalization of the electromagnetic fields are given (up to a scale factor) 
in terms of the 
time-averaged energy density, Eq.(\ref{effective Hamiltonian}):
\begin{equation}
\sum_{\mu \nu} \left(\bm u_{\mu})^*, \tilde{\bm B}^{-1}(\omega_{\mu}) 
\bm u_{\nu} \right) = \frac{1}{\omega_{\mu}} \delta_{\mu \nu} .
\end{equation}
Finally, the matrix $\tilde{\bm B}^{-1}$ and \textit{not} $\bm B^{-1}$ 
enters the expression for the Berry connection, since it also defines the 
normalization of our states.  

\section{\label{app2}Numerical algorithms for bandstructure calculations}
In this section, we shall describe our formulation of the photonic
bandstructure problem which has been used in the computations of the
edge mode spectra.  

Since we always neglect absorption/emmision and other non-linear processes 
of light (i.e. we work within an approximation of photon number conservation), 
we seek a real-space Hamiltonian formulation of the bandstructure problem.  
A real-space method is desirable over existing Fourier-space methods for our
purposes; the modes we are particularly interested in are obtained in 
domain wall configurations of the Faraday ``mass term'' as a function of position, 
and it is most simple and suitable to work within a real space formulation.   

In the numerical implementation of a Hamiltonian formulation, 
we shall treat the 
continuum flux densities $\langle \bm{y}| = (\bm{D}, \bm{B})$ 
rather than $\langle \bm r|  = (\bm{E}, \bm{H})$ as our fundamental
dynamical variables.  The former set obey the source-free Gauss' relations:
\begin{equation}
\nabla \cdot | \bm{y} \rangle = 0 .
\end{equation}

The Hamiltonian of our system is given by the following quadratic form:
\begin{equation}
\mathcal{H} = \frac{1}{2} \left(\bm{D},\epsilon^{-1} \bm{D} \right) + 
\frac{1}{2} \left(\bm{B},\mu^{-1} \bm{B} \right) .
\end{equation}

Furthermore, the propagating solutions of Maxwell's equations 
require the fields to be coupled in non-canonical Poisson bracket relations:
\begin{equation}
\lbrace D^a(x),B^b(x') \rbrace = \epsilon^{abc} \nabla_c \delta^3 (x-x')
\end{equation}

The two sets of fields are related by $|\bm{y} \rangle = \bm{B} |\bm r \rangle$, 
where $\bm{B}$ is the matrix of constitutive relations introduced in
section II.  
The source free Maxwell equations are slight variants of the ones described 
in section II.  Written as a generalized eigenmode problem of form
\begin{equation}
\bm{A} \tilde{\bm{B}} | \bm{y} \rangle = \omega |\bm{y} \rangle .
\label{genevproblem}
\end{equation}

The matrix $\bm{A}$ is the imaginary anti-symmetric matrix introduced in
section II, and $\tilde{\bm{B}} = \bm{B}^{-1}$ is a positive-definite
Hermitian matrix.  The eigenmode problem here is formally analogous to the 
problem of a non-canonical harmonic oscillator with Hamiltonian
\begin{equation}
\mathcal{H} = \frac{1}{2}\sum_{ij} \tilde{B}_{ij} y_i y_j ,
\end{equation}
and Poisson brackets
\begin{equation}
\lbrace y_i,y_j \rbrace = A_{ij} .
\end{equation}

Since $\bm{A}$ is imaginary and anti-symmetric, its eigenvalues are either 
zero, or come in pairs with opposite sign.  It is the presence of zero modes
which prevents a canonical treatment of the problem.  In the Maxwell problem, 
one third of the $\bm{A}$ matrix eigenvalues are zero modes.  
\begin{figure}[htbp]
\begin{center}
\psfig{file=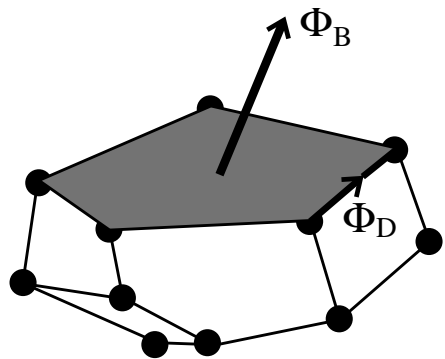}
\end{center}
\caption{The generic discretization scheme for the photon Band structure problem.  
Space is broken up into polyhedra.  Local electric energy density is defined at 
the vertices of each polyhedron, and the electric fluxes, defined on the edges 
of the polyhedron, connect two electric energy sites.  The volume of the polyhedron 
is associated with local magnetic energy density, and magnetic fluxes ``live'' on 
the faces of the polyhedron.  The scheme here has electric-magnetic duality 
in that a dual polyhedron can be defined on the vertices of which magnetic 
energy density defined, etc.  The scheme here is inspired by lattice QED, 
which ensures the correct long wavelength photon dispersion; the 
only difference here is absence of sources.}
\end{figure}

In the spatial discretization of this problem, we divide space into 
polyhedral cells, whose vertices contain the local electrical energy
density as well as the inverse permittivity tensor $\epsilon^{-1}_{ij}(\bm r)$.  
The electrical fluxes, $\Phi_D$ are defined on the edges of
the polyhedron, while the magnetic fluxes, $\Phi_B$ are associated with
the faces of each cell.  Finally, magnetic energy and the local
inverse permeability tensor $\mu^{-1}_{ij}(\bm r)$ are defined
on the centers of each polyhedron.  

This discretization scheme preserves the self-duality of the source-free
Maxwell equations in three dimensions; for each such electric polyhedron 
described above, there is a dual magnetic polyhedron whose faces correspond 
to the edges of the electric polyhedron, and whose center corresponds to
the vertices of the electric polyhedron.  

\begin{figure}[htbp]
\begin{center}
\psfig{file=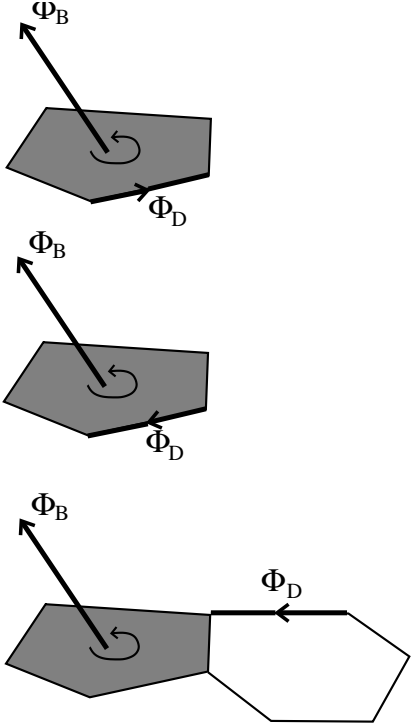}
\end{center}
\caption{The discretized form of the $\bm{A}$, which contains the Poisson
bracket relations of the fluxes.  Shown here are example configurations of 
$\lbrace \Phi^D_i,\Phi^B_j \rbrace = +i$ (top), $-i$ (middle), and $0$ (bottom).  }
\label{Amatrix}
\end{figure}

The discretized form the $\bm{A}$ matrix couples electric fluxes to 
magnetic ones, and vice-versa.  The coupling is (see Fig. \ref{Amatrix})
\begin{equation}
A_{ij}^{DB} = \lbrace \Phi_i^D,\Phi_j^B \rbrace = 0, \pm i .
\end{equation}

\begin{figure}[htbp]
\begin{center}
\psfig{file=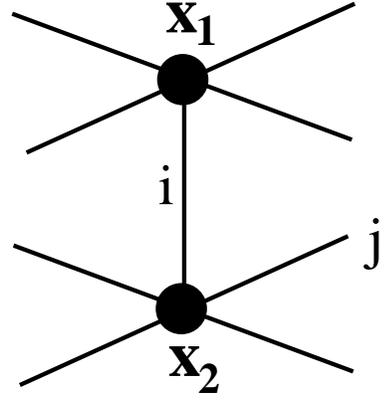}
\end{center}
\caption{The discretized form of the $\bm{B}$, which contains the contains
the geometric as well as the dynamics information.  It couples fluxes of the 
same type only, and allows for anisotropy in the constitutive relations.}
\label{Bmatrix}
\end{figure}

The $\bm{B}$ matrix couples fluxes of the same type, and depends on 
the geometry of the polyhedra used to discretize space.  For the case of 
a simple cubic discretization, and for the electric fluxes (see Fig. \ref{Bmatrix}), 
\begin{equation}
B_{ii} = \frac{1}{2} \left(\epsilon_{ii}^{-1}(\bm r_1) + \epsilon_{ii}^{-1}(\bm r_2) \right)
\end{equation}    
\begin{equation}
B_{ij} = \frac{1}{4} \epsilon^{-1}_{ji}(\bm r_2) .
\end{equation}

Identical relations involving the inverse permeability tensor are constructed
for the magnetic fluxes.  With the present formulation, the complete Hamiltonian of
the system is expressed as a sum of local terms, $\mathcal{H} = \sum_n h(\bm r_n)$, with
\begin{equation}
h(\bm r_n) = \sum_{ij}B_{ij}(\bm r_n) y_i y_j .
\end{equation}
Using this method, we can handle the case where the constitutive
relations have generalized anisotropy, and vary in space.  

We have made use of a simple cubic discretization of this general algorithm 
(although similar implementations using BCC and FCC lattices have given the
correct count of the long-wavelength photon modes) to compute photonic
band structures.  The fluxes $\Phi^D$ and $\Phi^B$ are made to obey the 
generalized Bloch boundary conditions
\begin{equation}
\Phi^{\sigma}(\bm{x} + \bm{R}) = e^{i\bm{k} \cdot \bm{R}} \Phi^{\sigma}(\bm{x}), \sigma 
= D,B ,
\end{equation}
where $\bm{R}$ is a lattice translation of the particular photonic crystal
under consideration, and $\bm{k}$ is a Bloch vector in the first Brillouin
zone.  We compute the bandstructure of the system by varying the Bloch vector
in the boundary conditions, which introduces Bloch phase factors into a few off-diagonal
elements of the $\bm{A}$ and $\bm{B}$ matrices and give rise to the
band dispersions.  

Both the $\bm{A}$ and $\bm{B}$ matrix are sufficiently sparse and are
stored as matrix-vector multipliers and are
treated using a Lanczos algorithm.  
However, due to the significantly large number of zero modes, 
a conventional Lanczos treatment of Eq.(\ref{genevproblem}) would not converge.  
The Lanczos adaptation for the 
Photonic problem is done by modifying the $\bm{A}$ matrix to
\begin{equation}
\bm{A} \rightarrow \bm{A}' = \bm{A} \bm{B} \bm{A} - 2 \omega_0 \bm{A} ,
\end{equation}
\begin{equation}
\bm{A}' \bm{B} |\bm{y} \rangle = \omega (\omega - 2\omega_0 ) | \bm{y} \rangle .
\end{equation}

Here, $\omega_0$ is the lowest eigenvalue, and the low lying (negative) 
eigenvalues of the modified problem are now the physically relevant ones 
which are easily found with the Lanczos implementation.  The dimensions of the
matrices are $d = 6N$, where N is the number of points used to discretize the
constitutive relations.  We have found system sizes up to $10^6$ to be accessible
within this approach.  Furthermore, local polarization and magnetization modes
can be added to the algorithm.  

\section{\label{app3}Derivation of the Dirac Point Splitting}
In this section, we derive a general expression for the frequency splitting 
at the Dirac point caused by inversion or time-reversal symmetry 
breaking perturbations.  We will use ``Bra-ket'' notation to represent 
our eigenvectors instead of writing equations for each component.  
We suppose that we know the exact eigenstates of the problem
\begin{equation}
\bm A | \bm u_0 \rangle = \omega_D \bm B_0^{-1} | \bm u_0 \rangle ,
\end{equation}
and that the solutions are two fold degenerate at the Dirac point, as for 
example, in the numerical examples we have considered.  
Now add a perturbation in the constitutive relations:
\begin{equation}
\bm{B}^{-1} = \bm{B}^{-1}_0 + \lambda \bm{B}^{-1}_1 .
\end{equation}

This term represents our inversion or time-reversal breaking perturbation.  
To find the splitting of the Dirac point (our ``Dirac mass''), we solve
the modified problem
\begin{equation}
\bm{A} | \bm{u}\rangle = \omega \left(\bm{B}^{-1}_0 + \lambda \bm{B}^{-1}_1 \right)
| \bm{u} \rangle .
\end{equation}

Since the $\bm{B}^{-1}_0$ matrix is positive-definite, it has a 
well defined positive-definite inverse square root matrix $\bm{B}_0^{1/2}$, 
and we can rewrite the unperturbed problem in the form of a
conventional Hermitian eigenvalue problem
\begin{equation}
\bm{B}_0^{1/2} \bm{A} \bm{B}_0^{1/2} | \bm{z}_0 \rangle = \omega_D | \bm{z}_0 \rangle ,
\end{equation}
where
\begin{equation}
| \bm{z}_0 \rangle = \bm{B}_0^{-1/2} | \bm{u}_0 \rangle .
\end{equation}

The new eigenvalue problem with the symmetry breaking terms is
\begin{eqnarray*}
\bm{A} | \bm{u} \rangle &=& \omega \left( \bm{B}^{-1}_0 + 
\lambda \bm{B}^{-1}_1 \right)| \bm{u} \rangle \\
&=& \omega \bm{B}_0^{-1/2} \left( 1 + \lambda \bm{B}_0^{1/2} 
\bm{B}^{-1}_1 \bm{B}_0^{1/2} \right) \bm{B}_0^{-1/2} | \bm{u} \rangle , 
\end{eqnarray*}
which subsequently is rewritten in the canonical form as
\begin{equation}
\bm{B}_0^{1/2} \left(\bm{A} -  \lambda \omega \bm{B}^{-1}_1\right) \bm{B}_0^{1/2} 
| \bm{z} \rangle = \omega | \bm{z} \rangle ,
\end{equation}
where $ | \bm{z} \rangle = \bm{B}_0^{-1/2} | \bm{u} \rangle$.  The 
correction to the spectrum to first order in perturbation theory in the
eigenvalue problem above is then
\begin{eqnarray*}
\delta \omega &=& -\omega_D \lambda \langle \bm{z}_0 | \bm{B}_0^{1/2} 
\bm{B}^{-1}_1 \bm{B}_0^{1/2} | \bm{z}_0 \rangle  \\
&=& -\omega_D  \lambda \frac{\langle \bm{u}_0 | \bm{B}^{-1}_1 | 
\bm{u}_0 \rangle}{\langle \bm{u}_0 | \bm{B}^{-1}_0 | \bm{u}_0 \rangle} .
\end{eqnarray*}

We have restored the normalization factor for the state 
$|\bm{u}_0 \rangle$ in the last line above.  
Thus, our main result here is a general expression for
the splitting of the Dirac point frequency spectrum, given by the 
dimensionless quantity
\begin{equation}
\frac{\delta \omega}{\omega_D} = -\lambda \frac{\langle \bm{u}_0 | \bm{B}^{-1}_1 | 
\bm{u}_0 \rangle}{\langle \bm{u}_0 | \bm{B}^{-1}_0 | \bm{u}_0 \rangle } .
\end{equation}
  
\end{appendix}


\begin{thebibliography}{999}

\bibitem{joannopoulos}
J. D. Joannopoulos et.al.  \emph{Photonic Crystals: Molding the Flow of Light}. 
Princeton University Press, 1995.

\bibitem{left-handed}
R. Marques, J. Martel, F. Mesa, and F. Medina, Phys. Rev. Lett. 
\textbf{89} 183901 (2002).

\bibitem{lincoln logs}
Shawn-Yu Lin, Edmund Chow, Vince Hietala, Pierre R. Villeneuve, 
Science \textbf{282} 274 (1998).

\bibitem{sundaram niu}
Sundaram and Q. Niu, Phys. Rev. B. 

\bibitem{simon}
B. Simon, Phys. Rev. Lett. \textbf{51}, 2167 (1983).


\bibitem{berry}
M. V. Berry, Proc. Roy. Soc. London A \textbf{392}, 45 (1984).


\bibitem{nagaosa}
M. Onoda, S. Murakami, and N. Nagaosa, Phys. Rev. Lett.  93, 083901 (2004)


\bibitem{raghu}
F. D. M. Haldane and S. Raghu, Cond-mat/0503588.  



\bibitem{ahe}
Wei-Li Lee, Satoshi Watauchi, V.L. Miller, R. J. Cava, and N. P. Ong, 
Science \textbf{303}, 1647 (2004).

\bibitem{qhe}
K. von Klitzing, G. Dorda, and M. Pepper, Phys. Rev. Lett. \textbf{45}, 494 (1980)  


\bibitem{laughlin} R. B. Laughlin, Phys. Rev. Lett. \textbf{50}, 1395 (1983). 

\bibitem{hofstadter}
D. R. Hofstadter, Phys. Rev. B \textbf{14}, 2239 (1976).

\bibitem{niu 2}
M. C. Chang and Q. Niu, Phys. Rev. B \textbf{53} 7010 (1996).

\bibitem{tknn}
D. Thouless, M. Kohmoto, M. Nightingale, M. den Nijs, Phys. Rev. Lett. \textbf{49}, 405, (1982). 


\bibitem{graphene}
F. D. M. Haldane, Phys. Rev. Lett. \textbf{61} 2015, (1988).


\bibitem{nagaosa1} 
M. Onoda, N. Nagaosa, Phys. Rev. Lett. \textbf{90} 206601 (2003).

\bibitem{kanemele} 
C. L. Kane, E. J. Mele, Phys. Rev. Lett. \textbf{95}, 146802 (2005)
C. L. Kane, E. J. Mele, Phys. Rev. Lett. \textbf{95}, 226801 (2005)


\bibitem{karplusluttinger} 
R. Karplus and J. M. Luttinger, Phys. Rev. \textbf{95}, 1154 (1954)

\bibitem{kohmoto}
M. Kohmoto, Ann. Phys. (NY) \textbf{160} 343 (1985).  

\bibitem{faraday}
R. C. Booth and E. A. D. White.  J. Phys. D: Appl. Phys., \textbf{17} 579-587 (1984).

\bibitem{landauIII}
L. D. Landau and L. P. Lifshitz.  \emph{Quantum Mechanics: Non-Relativistic Theory}.

\bibitem{maradudin} 
M. Plihal and A. A. Maradudin, Phys. Rev. B 44 16 8565 (1991).


\bibitem{thouless}
D. J. Thouless, \emph{Topological Quantum Number in Nonrelativistic Physics}.  
World Scientific, 1998.





\bibitem{footnote1}
We shall temporarily use the inverse permittivity in this subsection since 
it shall enable us to derive a simple differential for the scalar magnetic 
field of the TE modes.  

\end{thebibliography}
\end{document}